\documentclass[aps,preprintnumbers, superscriptaddress, amsmath, showpacs]{revtex4}
\usepackage{psfrag}
\usepackage{epsfig}
\usepackage{amsfonts} % para incluir fonts y
\usepackage{amssymb}  % simbolos de la AMS
\usepackage{graphicx} % standard LaTeX graphics tool
\usepackage{amsmath}  % for including eps-figure files
\usepackage{amsmath}  % paquete matemÃ¯Â¿Â½tico

\usepackage[latin1]{inputenc}
\usepackage{color}
\usepackage{suffix}
\usepackage{mathtools}

\newcommand{\slsh}[1]{{\not \! #1}}
\newcommand{\Eq}[1]{{Eq.~({\ref{#1}})}}
\newcommand{\bea}{\begin{eqnarray}}
\newcommand{\eea}{\end{eqnarray}}
\newcommand{\beas}{\begin{eqnarray*}}
\newcommand{\eeas}{\end{eqnarray*}}
\newcommand{\sumint}{\sum\!\!\!\!\!\!\!\!\int}

\DeclarePairedDelimiterX\MeijerM[3]{\lparen}{\rparen}%
{\begin{smallmatrix}#1 \\ #2\end{smallmatrix}\delimsize\vert\,#3}

\newcommand\MeijerG[8][]{%
  G^{\,#2,#3}_{#4,#5}\MeijerM[#1]{#6}{#7}{#8}}

\WithSuffix\newcommand\MeijerG*[7]{%
  G^{\,#1,#2}_{#3,#4}\MeijerM*{#5}{#6}{#7}}

\def\Title#1{\begin{center} {\Large {\bf #1} } \end{center}}

\begin{document}

\Title{(In-)Significance of the Anomalous Magnetic Moment of Charged Fermions for the Equation of State of a Magnetized and Dense Medium}

\author{E. J. Ferrer, V de la Incera}
\affiliation{Department of Physics, University of Texas at El Paso, El Paso, TX 79968, USA}
 \author{D. Manreza Paret}
\affiliation{Departamento de Fisica General, Facultad de Fisica, Universidad de la Habana, La Habana, 10400, Cuba}
\author{A. P\'{e}rez Mart\'{\i}nez}
\affiliation{Instituto de Cibern\'{e}tica, Matem\'{a}tica y F\'{\i}sica (ICIMAF),
 La Habana, 10400, Cuba}
 \author{A. Sanchez}
\affiliation{Facultad de Ciencias, Universidad Nacional Aut\' onoma de M\' exico, Apartado Postal 50-542, M\'exico Distrito Federal 04510, M\'exico.}

\begin{abstract}

We investigate the effects of the anomalous magnetic moment (AMM) in the equation of state (EoS) of a
system of charged fermions at finite density in the presence of a magnetic field. In the region of strong magnetic fields ($eB>m^2$) the
AMM is found from the one-loop fermion self-energy. In contrast to the weak-field AMM found by
Schwinger, in the strong magnetic field region the AMM depends on the Landau level
and decreases with it. The effects of the AMM in the EoS of a dense medium are investigated at strong and weak fields using the appropriate AMM expression for each case. In contrast with what has been reported in other works, we find that the AMM of charged fermions makes no significant contribution to the EoS at any field value.\end{abstract}


%=======================================================================================
%\pacs{26.30.+k, 91.65.Dt, 98.80.Ft}% PACS,  the Physics and Astronomy

\maketitle

\section{Introduction}

The fact that strong magnetic fields populate the vast majority of the astrophysical compact objects and that they can significantly affect several properties of the star have served as motivation for many works focused on the study of the Equation of State (EoS) of magnetized systems of fermions and their astrophysical implications \cite{MNS}-\cite{EOS-Ferrer}.

In the presence of a magnetic field $B$ the dispersion relation of charged fermions takes the form $E=\sqrt{p^2_3+2eBl+m^2}$, exhibiting the Landau quantization of the cyclotron frequencies characterized by the Landau-level number $l=0,1,2,...$  \cite{Landau}. A magnetic field also affects the density of states which now becomes proportional to the field, so the three-momentum integrals change as
\begin{equation}
2\int\frac{d^3p}{(2\pi)^3}\rightarrow\sum_l g(l)\frac{eB}{(2\pi)^2}\int dp_z
\end{equation}\label{density_state}
The factor $g(l)=[2-(\delta_{l0})]$ takes into account the double spin degeneracy of all the Landau levels except $l=0$. In addition, a magnetic field breaks the rotational
SO(3) symmetry, giving rise to an anisotropy in the energy-momentum tensor \cite{Landau-Lifshitz} and producing a pressure splitting in two distinguishable components, one along the field (the longitudinal pressure) and another in the perpendicular direction (the transverse pressure). As a consequence, a system of fermions in a constant and uniform magnetic field exhibits an anisotropic EoS \cite{Canuto}-\cite{EOS-Ferrer}.

Besides modifiying the one-particle Dirac Hamiltonian, a
magnetic field can also affect the radiative corrections of the
fermion self-energy because it introduces an
additional tensor $F_{\mu\nu}$ in the theory that gives rise to new independent structures like $\frac{1}{2}{\scriptstyle \mathcal{T}} \sigma_{\mu\nu}F^{\mu\nu}$, with $ \sigma_{\mu\nu}= \frac{i}{2}[\gamma_\mu,\gamma_\nu]$. This new term corresponds to the coupling between the field and the fermion anomalous magnetic moment (AMM) ${\scriptstyle \mathcal{T}}$ \cite{Schwinger}, which in general can be  a function of the magnetic field. It induces a Zeeman splitting in the fermion dispersion that removes the spin degeneracy \cite{AMM-Old}, and the following change in the density of states
\begin{equation}
\sum_l g(l,\sigma)\frac{eB}{(2\pi)^2}\int dp_z, \qquad g(l,\sigma)=\delta_{l0}+(1-\delta_{l0})\sum_{\sigma=\pm 1}
\end{equation}\label{density-state-2}
with the spin projections $\sigma=\pm1$.

As shown by Schwinger \cite{Schwinger} many years ago, at weak fields ($eB\ll m^2$), one can make an expansion of the fermion self-energy in powers of the magnetic field and find that the leading contribution proportional to $\sigma_{\mu\nu}$ is linear in the field and given (in natural units) by ${\scriptstyle \mathcal{T}} B$ with ${\scriptstyle \mathcal{T}}=(\alpha/2\pi)\mu_B$, where $\mu_B=e/2m$ is the Bohr magneton. In this approximation, ${\scriptstyle \mathcal{T}}$ is simply independent of the magnetic field. At strong field, however, the coefficient of the $\sigma_{\mu\nu}$ structure varies as a square logarithm of the field  \cite{Jancovici} and hence cannot be expanded in powers of the magnetic field. At finite temperature and/or density, the concept of weak field needs to be revisited, as the field may be strong with respect to one of the scales, but weak with respect to the others.

Charged and massive fermions always possess magnetic moment which in principle can produce interesting physical effects via the modification of the self-energy. Moreover, massless charged fermions in the presence of a magnetic field can acquire a dynamical magnetic moment \cite{Vivian}-\cite{AMM-NJL} through the phenomenon of magnetic catalysis of chiral symmetry breaking (MC$\chi$SB) \cite{Igor}. The mechanism responsible for this effect is related to the dimensional reduction of the infrared dynamics of the particles in the lowest Landau level (LLL). Such a reduction favors the formation of a chiral condensate because there is no energy gap between the infrared fermions in the LLL and the LLL antiparticles in the Dirac sea. This effect has been actively investigated for the last two decades \cite{Vivian}-\cite{Incera}. In the original studies of the MC$\chi$SB phenomenon \cite{Igor}-\cite{Incera}, the catalyzed chiral condensate was assumed to generate only a fermion dynamical mass. Recently, however, it has been shown that in QED \cite{Vivian}-\cite{VyE}, as well as in quark systems with \cite{Bo-CS} and without \cite{AMM-NJL} finite density, the MC$\chi$SB inevitably leads  to a dynamical AMM together with a dynamical fermion mass.  In massless QED, the dynamical AMM gives rise to a non-perturbative Lande g-factor, a Bohr magneton proportional to the inverse of the dynamical mass, and to the realization of a non-perturbative Zeeman effect  \cite{Vivian}-\cite{VyE}.

The AMM term in the Hamiltonian changes the energy spectrum of the fermions and can affect in principle the properties of the system. Notice that certain neutral particles which, like the neutron, are composed by charged particles (charged quarks in this case), can also have nonzero AMM. The effects of the nucleons' and quarks' AMM on the statistics of magnetized matter have been discussed in many works \cite{Broderick, EoS-AMM}. The AMM has been linked among other strong-field effects to stiffening the EoS in magnetized stars and to a dramatic variation of the particle fraction, which at very high magnetic fields would lead, for example, to pure neutron matter (in the papers of Ref. \cite{Broderick, EoS-AMM} one or both of these findings are discussed). However, when investigating the effects of the AMM in any physical process we should be careful in using the analytic expression of the AMM that is consistent with the magnetic-field strength under consideration. In particular, as we will discuss in detail below, considering a linear-in-B approach, which is the approximation used throughout all the Refs. \cite{Broderick, EoS-AMM}, is only consistent in the region of weak magnetic fields where a large number of Landau levels are occupied because $\sqrt{eB}$ is smaller than all the other energy scales like mass, temperature, and chemical potential.

The critical field, below which the weak approximation is reliable then depends on the content of the stellar matter, the temperature and the density. If one can ignore $T$ and $\mu$, the critical field separating the weak and strong field regions is determined by the particle mass. For each particle species, the critical field can then be obtained by equating the magnetic energy $\hbar \omega_c$, where $\omega_c=qB/mc$ is the cyclotron frequency in cgs units, to the corresponding rest-energy $mc^2$. The range of critical fields is then quite wide. For example, for electrons,  $B_c^{(e)}=4.4\times 10^{13}$ Gauss, for quark matter formed by $u$ and $d$ quarks with current masses $m_u=m_d=5$ MeV/$c^2$, it is $B_c^{(u,d)}=10^2B_c^{(e)}=4.4\times 10^{15}$ Gauss, for protons, whose mass is  $938$ MeV$/c^2$, one finds $B_c^{(p)}=1.6\times 10^{20}$ Gauss, while charged hyperons, which are much heavier, will have a critical field two orders of magnitude larger than protons'. At zero temperature and density, a field larger than the critical one for that type of particle will constraint them to their LLL. In a system with different types of particles, a field may be strong, hence over the critical field, for some of them and weak, below the critical, for others, so care must be taken when using weak and strong field approximations to consider such subtleties.

In the present paper we are interested in revising the role of the AMM in the EoS of systems of charged fermions, under both
 weak and strong magnetic fields. This is a due task given that in several of the works that have studied the effects of the AMM in the thermodynamical properties one can point out several issues in the way the results have been obtained. One of these issues is that the strong-field region has been explored inconsistently considering Schwinger's result for the AMM of all the particles, thus ignoring the existence of different critical fields and the fact that the Schwinger's approximation for the AMM breaks down for fields of the order of or larger than the critical one,  as pointed out many years ago in Ref. \cite{Jancovici}. Second, when calculating the pressure, some works have ignored the existence of a pressure anisotropy \cite{Canuto}-\cite{EOS-Ferrer} in a strong magnetic field, so the results were obtained basically using a single pressure. Third, some papers neglected the contribution of the Maxwell magnetic pressure proportional to $B^2/2$ and claimed that the Schwinger AMM produced a significant contribution to the statistical quantities, but they concluded this by considering a region of fields where the Schwinger approximation not only breaks down, but the magnetic pressure can dominate the matter pressure and erase any possible effect of the AMM, as will be shown in this paper. Fourth, the pressure of the magnetized vacuum, that is, the contribution that does not depend on temperature or density, was also neglected at strong fields where it can be important. To clarify all these issues, we shall analyze, through analytical and numerical calculations, the significance of the AMM contribution to the main statistical quantities of the magnetized system in the weak and strong-field approximations, as well as to the EoS of dense systems with an interest for astrophysical applications. With this goal in mind, we shall investigate the weight of each of the participating contributions (Maxwell pressure, vacuum pressure, etc.) into the system EoS for different field values. As will be demonstrated, our thorough analysis leads us to conclude that, when working consistently, the quantum effect of the AMM of charged particles is negligibly small for the EoS of the magnetized system at both weak and strong fields.

The paper is organized as follow. In Sec. \ref{section2} we give the one-loop self-energy of a charged fermion system in the presence of a constant and uniform magnetic field using the Ritus's method \cite{Ritus:1978cj}, and find the AMM analytical expression for the different LL's in the strong-field limit. In Sec. \ref{section3}, it is calculated the one-loop thermodynamical potential depending on the AMM in the strong-field approximation. The result is given as the sum of the renormalized vacuum contribution, the contribution at zero temperature and finite density and the thermal contribution. In Sec. \ref{section4}, for the sake of completeness, we calculate the AMM in the weak-field approximation using a combination of Ritus eigenfunction method and proper-time representation.  In Sec. \ref{section5}, we calculate the one-loop thermodynamic potential in the weak-field approximation including the AMM found in that approximation. The renormalized vacuum contribution and the zero-temperature finite-density contribution are presented up to $\mathcal{O}((eB)^4)$ order. In Sec. \ref{section7}, we present the numerical results for the main thermodynamic quantities, which depend on the AMM, in the weak- and strong-magnetic-filed limits. There, we make a thoughtful analysis to determine the significance of the AMM for the EoS of strongly and weakly magnetized systems of charged fermions. Finally, in Sec. \ref{section8} we state our concluding remarks. We also include four Appendices. In Appendix A, we give details on the calculation of the thermodynamical potential in the strong-field approximation at $T\neq 0$ and $\mu\neq 0$. In Appendix B, we discuss some issues in the calculation of the effective potential at $B\neq 0$ in the Dittrich's approach. In Appendix C, we derive the Schwinger propagator at $B\neq 0$, starting from the Ritus's formalism. In Appendix D, the details of the calculation of the thermodynamic potential in the weak-field approximation at $T\neq 0$ and $\mu\neq 0$ are given.

\section{AMM in the strong-field approximation}\label{section2}

The radiative corrections to the magnetic moment of a charged particle in
the presence of a magnetic field can be found from the one-loop fermion self energy

\begin{equation}\label{SE-coordinate}
\Sigma (x,x')=-ie^2 \gamma^{\mu}G(x,x')\gamma^{\nu}D_{\mu
\nu}(x-x'),
\end{equation}
$G(x,x')$ denotes the fermion's propagator in the presence of a uniform and constant magnetic field and $D_{\mu
\nu}(x-x')$ is the photon propagator.

One can transform the self-energy to momentum space by using Ritus's approach

\begin{eqnarray}\label{P-Self-Energy}
\Sigma(p,p')= \int d^4xd^4y
\overline{\mathbb{E}}_{p}^{l}(x)\Sigma(x,y)\mathbb{E}_{p'}^{l}(y)=(2\pi)^4\widehat{\delta}^{(4)}(p-p')\Pi(l)\widetilde{\Sigma}^l
(\overline{p}),
\end{eqnarray}
Index $l$ denotes the Landau-level number; $\Pi(l) = \Delta(\mathrm{sgn}(eB))\delta^{l0}+I(1-\delta^{l0})$ is a projector that separates the LLL ($l=0$), with a single spin projection, from the rest ($l>0$) with two; $\widehat{\delta}^{(4)}(p-p')=\delta^{ll'} \delta(p_{0}-p'_{0}) \delta(p_{2}-p'_{2}) \delta(p_{3}-p'_{3})$. The Ritus eigenfuntions \cite{Ritus:1978cj} are given by
\begin{equation}\label{Ep}
 \mathbb{E}_{p}^{l}(x)=\sum_{\sigma=\pm1}E_{p}^{\sigma}(x)\Delta(\sigma),
\qquad \overline{\mathbb{E}}_{p}^{l}\equiv \gamma^{0}
(\mathbb{E}_{p}^{l})^{\dag}\gamma^{0}
\end{equation}
with
\begin{equation}
\Delta(\pm)=\frac{I\pm i\gamma^{1}\gamma^{2}}{2},
\label{Spin-projectors}
\end{equation}
are spin up ($+$) and down ($-$) projectors, and
\begin{eqnarray}\label{E-x}
E_{p}^{+}(x)=N_{l}e^{-i(p_{0}x^{0}+p_{2}x^{2}+p_{3}x^{3})}D_{l}(\rho),\qquad
\nonumber
\\
E_{p}^{-}(x)=N_{l-1}e^{-i(p_{0}x^{0}+p_{2}x^{2}+p_{3}x^{3})}D_{l-1}(\rho)
\end{eqnarray}
with normalization constant $N_{l}=(4\pi eB)^{1/4}/\sqrt{l!}$, and
$D_{l}(\rho)$  are the parabolic cylinder functions of argument
$\rho=\sqrt{2eB}(x_{1}-p_{2}/eB)$.

In momentum space the general structure of the self energy is \cite{Incera}
\begin{equation}\label{SE-structure}
\Sigma^{l}(\overline{p})
=Z_{\|}^{l}\overline{p}_{\|}^\mu\gamma_{\mu}^{\|}+Z_{\bot}^{l}\overline{p}_{\bot}^\mu\gamma_{\mu}^{\bot}+M_{l}I+i\mathcal{T}_{l}\gamma^{1}\gamma^{2},
\end{equation}
Notice the separation between parallel ${\overline{p}}_{\|}^\nu=(p^{0},0, 0,p^{3})$ and perpendicular ${\overline{p}}_{\bot}^\nu=(0,0, \sqrt{2eBl},0)$ components due to the spatial symmetry breaking in a magnetic field that only leaves intact the subgroup of rotations along the field direction. In (\ref{SE-structure}), $Z_{\|}^{l}$, $Z_{\bot}^{l}$ are the wave function's renormalization coefficients. The coefficients $M_{l}$ and $\mathcal{T}_l$ are respectively the radiative corrections to the mass and the magnetic moment. Each of them has to be determined as a solution of the Schwinger-Dyson (SD) equations of the theory at the given approximation.

In the one-loop approximation, the Schwinger-Dyson equation leads to an infinite set of couple equations that take the form  \cite{VyE, Vivian, AuroraDaryelVF}
\begin{eqnarray}
\Sigma^{l}(\overline{p})\Pi(l)&=&-ie^2(2eB)\Pi(l)\int\frac{d^4\widehat{q}}{(2\pi)^4}
\frac{e^{-\widehat{q}^2_\bot}}{\widehat{q}^2}[L_l+L_{l+1}+L_{l-1}], \quad l=0,1,2,....\label{SD-EqL}
\end{eqnarray}
with
\begin{eqnarray}
  L_l &=& \gamma_{\mu}^{\|}G^{l}(\overline{p-q})\gamma_{\mu}^{\|}, \nonumber\\
  L_{l\pm1} &=& \Delta(\pm)\gamma_{\mu}^{\bot}G^{l\pm1}(\overline{p-q})\gamma_{\mu}^{\bot}\Delta(\pm)
 \label{expresiones1}
\end{eqnarray}
and fermion propagator
\begin{equation}
G^l(\overline{p})=\dfrac{\overline{p}\cdot\gamma+m}{\overline{p}^2-m^2}\Pi(l),
\label{Fermion propagator}
\end{equation}
Here, we introduced the notation $\widehat{q}_\mu=q_\mu/\sqrt{2|eB|}$, $\overline{p}_\mu = (p^{0},0,-\sqrt{2|eB|l},p^{3})$ and $(\overline{p-q})_\mu = (p^{0}-q^{0},0,-\sqrt{2|eB|l},p^{3}-q^{3})$. Henceforth, we assume $eB>0$. As it will become clear below, the representation (\ref{SD-EqL}) of the self-energy is particularly convenient for strong-field calculations.

From Eqs. (\ref{SE-structure}) and  (\ref{SD-EqL}), we can extract the equations for the AMM at each LL. In Euclidean space they are
\begin{eqnarray}
E_0=(M_0+\mathcal{T}_0)=e^2m(4eB)\int\frac{d^4\widehat{q}}{(2\pi)^4}
\frac{e^{-\widehat{q}^2_\bot}}{\widehat{q}^2}\left(\frac{1}{(\overline{p-q})^2_0+m^2} +\frac{1}{(\overline{p-q})^2_{1}+m^2}\right ),\label{LLLMT}
\end{eqnarray}

\begin{eqnarray}
  \mathcal{T}_{l} &=&-e^2m(2eB)\int\frac{d^4\widehat{q}}{(2\pi)^4}\frac{e^{-\widehat{q}^2_\bot}}{\widehat{q}^2}.
  \left[\frac{1}{(\overline{p-q})^2_{l+1}+m^2}-\frac{1}{(\overline{p-q})^2_{l-1}+m^2} \right], \quad l\geq 1 \label{Tl}
\end{eqnarray}
Eq. (\ref{LLLMT}) reflects the single spin orientation of the fermions in the LLL ($l=0$) and hence the impossibility of determining $M_0$ and $\mathcal{T}_0$ independently \cite{VyE, Vivian}. Thus, $E_0$ cannot be interpreted as an AMM term, but as the radiative correction to the rest-energy of the LLL particles. Notice, that $E_0$ will not produce any Zeeman splitting in the modes of the LLL quasiparticles.

In the infrared limit $p_0=0, p_3\rightarrow 0$, and considering the strong-field approximation, the leading contributions to (\ref{LLLMT}) and (\ref{Tl}) for $l=1$ are respectively given by
\begin{equation}
  E_0=M_{0}+\mathcal{T}_{0} \simeq \frac{e^2m}{8\pi^3}\int d\widehat{q}_{\|}^2d\widehat{q}_{\bot}^2\frac{e^{-\widehat{q}^2_\bot}}{\widehat{q}^2}  \frac{1}{\widehat{q}_{\|}^2+\widehat{m}^2} =m\frac{\alpha}{4\pi}\ln^2({m}^2/2eB),
 \label{LLLMT0}
\end{equation}

\begin{equation}
  \mathcal{T}_{1} \simeq \frac{e^2m}{16\pi^3}\int d\widehat{q}_{\|}^2d\widehat{q}_{\bot}^2\frac{e^{-\widehat{q}^2_\bot}}{\widehat{q}^2}  \frac{1}{\widehat{q}_{\|}^2+\widehat{m}^2}=m\frac{\alpha}{8\pi}\ln^2({m}^2/2eB)
,\label{T1}
\end{equation}
Note that the leading contribution in (\ref{T1}) comes from the spin-down particles in the first LL (the term ${l-1}$ in (\ref{Tl}) for $l=1$), since $\sqrt{2eBl}$ acts, for $l\geq 1$, as a suppressing factor in the denominator of the fermion propagator. The result (\ref{LLLMT0}) coincides with that obtained many years ago in Ref. \cite{Jancovici} using a different method. As in massless-QED \cite{VyE, Vivian},  the relation $\mathcal{T}_1=E_0/2$ is satisfied here too.

For the remaining $\mathcal{T}_l$, $l>1$, we have
\begin{eqnarray}\label{strong-LL2}
  \mathcal{T}_{l} =&-&\frac{\alpha m}{16\pi^2}\,e^{-\hat{M}_{l+1}^2}\left\{
  -\gamma\Gamma[0,-\hat{M}_{l+1}^2]+e^2\gamma\Gamma[0,-\hat{M}_{l-1}^2]\right.\nonumber\\
  &-&i\pi e^2\ln\hat{M}_{l-1}^2- e^2 E_i(\hat{M}_{l-1}^2)\ln\hat{M}_{l-1}^2+i\pi\ln\hat{M}_{l+1}^2+E_i(\hat{M}_{l+1}^2)\ln\hat{M}_{l+1}^2\nonumber\\
  &-&\left.\MeijerG*{3}{0}{2}{3}{1, 1}{0, 0, 0}{-\hat{M}_{l+1}^2}+e^2\MeijerG*{3}{0}{2}{3}{1, 1}{0, 0, 0}{-\hat{M}_{l-1}^2}
  \right\},
\end{eqnarray}
with $\hat{M}^2_{l\pm1}=\hat{m}^2+(l\pm1)$, $\gamma\simeq0.577216$ is the Euler's constant, $E_i[z]$ denoting the exponential integral function, $\Gamma[0,z]$ is the incomplete gamma function and $\MeijerG*{m}{n}{p}{q}{a_1, \dots, a_p}{b_1, \dots, b_q}{z}$ the Meijer G-function~\cite{Gradshteyn}.

\begin{figure}[!ht]
\begin{center}
\includegraphics[width=12cm,height=10cm]{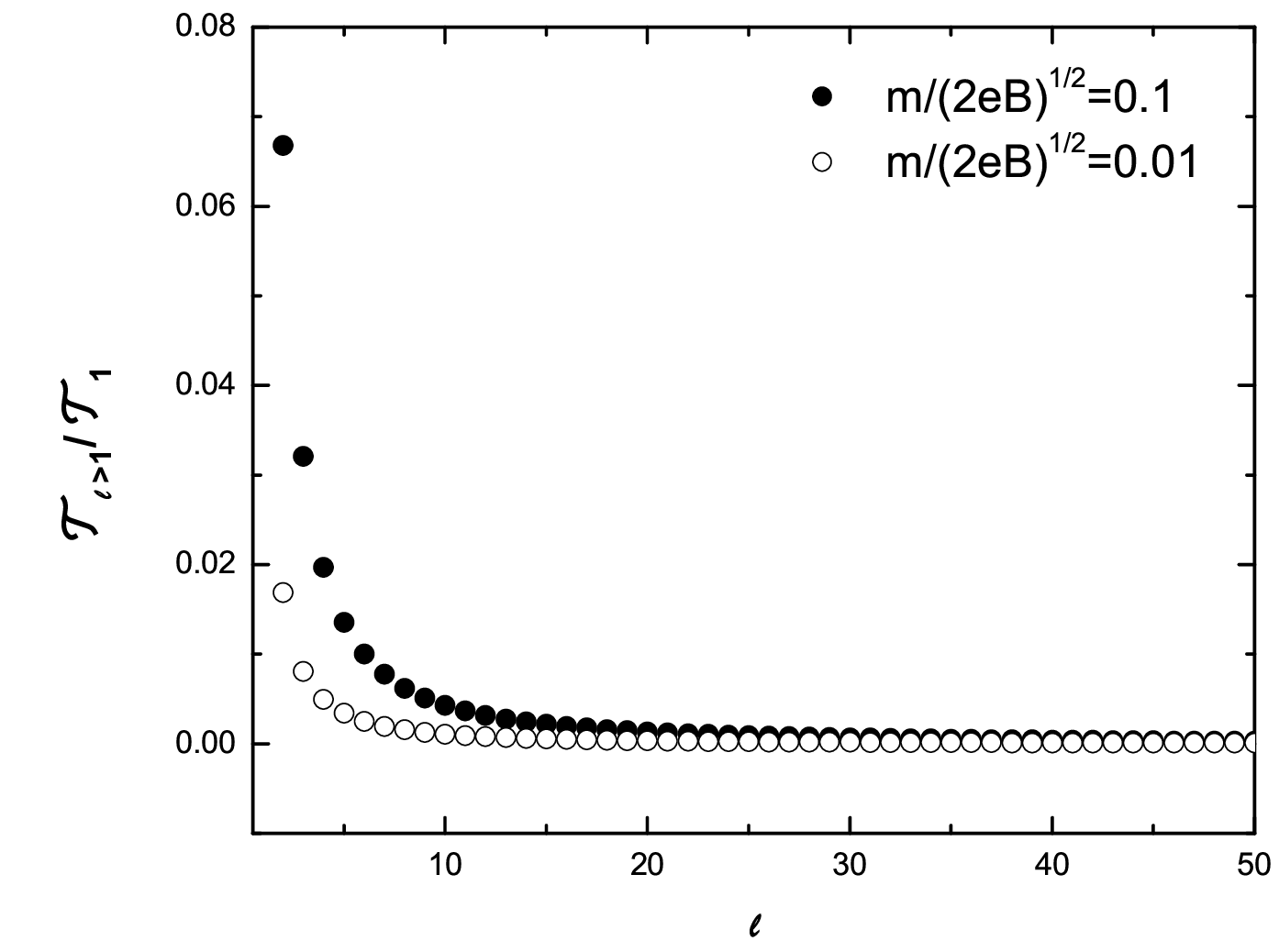}
\caption{\footnotesize \emph{Comparison between the AMMs $\mathcal{T}_{l}$ and $\mathcal{T}_{1}$ in the strong field region for Landau levels $l>1$}.The plot shows a sharp decrease of the ratio $\mathcal{T}_{l>1}/\mathcal{T}_{1}$ with increasing Landau levels for two field values. The largest values of $\mathcal{T}_{l>1}$ occur for the lowest ($l>1$)-value, but they are still two orders of magnitude smaller than $\mathcal{T}_{1}$. The stronger the field, the smaller the values of $\mathcal{T}_{l>1}$ for the same Landau level, and the quicker they approach to their asymptotic negligible value.} \label{fig1}
\end{center}
\end{figure}

In Fig.1 we show how the AMM's for $l>1$ decrease with respect to $\mathcal{T}_1$ as the LL increases.
Notice that the AMM at strong field is relevant only for the first Landau level, where it grows as the square logarithm of the field. As seen in Fig.\ref{fig1}, already in the second LL the AMM decreases in two orders with respect to its value at $l=1$, $\mathcal{T}_2/\mathcal{T}_1\sim0.0668$ for $\hat{m}=0.1$.

One can explicitly see from (\ref{LLLMT0})-(\ref{strong-LL2}) that, in contrast to the AMM found by  Schwinger \cite{Schwinger} in the weak-field limit, which was the same for all LL's and had a linear dependence with the field, at strong field the AMM does not depend linearly on the field and is different for each LL. Clearly, using the Schwinger AMM in the strong-field region would be totally inconsistent and care should be taken not to draw any physical conclusions obtained with such a wrong approach.

\section{Thermodynamic Potential with AMM in the Strong-Field Region}\label{section3}

To investigate the effects of the AMM in the EoS we consider an effective theory on which the fermion propagator is dressed by the one-loop fermion self-energy in the magnetic field, which depends on the AMM. The lack of Zeeman splitting (see Eq. (2)) in the LLL separates the propagator in the LLL from those in the rest of the levels, so the dressed propagators take the form\begin{equation}\label{Green-F-0}
G^{-1}_0(\overline{p})=(\overline{p}_\|\cdot\gamma^\|-m) \Delta(+)
\end{equation}
and
\begin{equation}\label{Green-F-L}
G^{-1}_l(\overline{p})= \overline{p}\cdot\gamma-m
      -i\mathcal{T}_l\gamma^1\gamma^2
\end{equation}
with ${\overline{p}}=(ip^{4},0, \sqrt{2|eB|l},p^{3})$ for $l=0,1,2,...$ in Euclidean space. Notice that in (\ref{Green-F-0}) we do not include the correction $E_0$,  neither in (\ref{Green-F-L}) the one-loop corrections to the mass, as they are negligible compared to the renormalized mass at $B=0$, $m$. However, since $\mathcal{T}_l$ gives rise to a new Lorentz structure, it is included in (\ref{Green-F-L}).

The fermion contribution to the thermodynamic potential of this effective theory is
\begin{equation}\label{Grand-Potential-4}
\Omega(B,\mu,T)= -\frac{eB}{\beta}\left[\sum_{p_4}\int\limits_{-\infty}^{\infty}\frac{dp_3}{(2\pi)^2} \ln \det G^{-1}_0(\overline{p}^*)+ \sum_{\sigma=\pm 1}\sum_{l=1}^{\infty}\sum_{p_4}\int\limits_{-\infty}^{\infty}\frac{dp_3}{(2\pi)^2} \ln \det G^{-1}_l(\overline{p}^*)\right]
\end{equation}
Here, $\beta = 1/T$  denotes the inverse temperature, $\mu$ the fermion chemical potential and ${\overline{p}}^*=(ip^{4}-\mu,0, \sqrt{2eBl},p^{3})$.

Performing the sum in Matsubara frequencies and calculating the determinants in Eq. (\ref{Grand-Potential-4}) we obtain
\begin{equation}\label{Omega-cosh}
\Omega(B,\mu,T)= -\frac{eB}{4\pi^2}\sum_{\eta\sigma l}\int dp_3\frac{1}{\beta} \ln \left[\cosh \frac{\beta}{2}(E_{\eta\sigma l}-\mu)\right],
\end{equation}
which can be rewritten as
\bea \label{omega-init}
   \Omega(B,\mu,T)&=&-\frac{1}{2} \frac{eB}{4\pi^2} \int_{-\infty}^{\infty}  dp_3  \sum_{\eta\sigma l} \vert E_{\eta\sigma l}-\mu\vert
    -\frac{eB}{4\pi^2} \int_{-\infty}^{\infty}  dp_3  \sum_{\eta\sigma l} \frac{1}{\beta} \ln(1+e^{-\beta\vert E_{\eta\sigma l}  -\mu\vert})
\eea
In these expressions the sum in the energies $E_{\eta\sigma l}$ include particles/antiparticles ($\eta=\pm$), up/down spin ($\sigma=\pm$), and Landau level $l$ indices.
For the LLL, only one spin projection contributes and the energy becomes

\bea \label{LLLmode}
   E_{\eta, 0}=\eta \sqrt{p_3^{2}+m^2} \quad \quad l=0, \quad \eta=\pm 1
\eea

For each $l\neq0$, the AMM separates the energies of up and down spin ($\sigma=\pm$) as
\bea
E_{\eta \sigma l}=\eta \sqrt{p_3^{2}+(\sqrt{2eBl+m^2}+\sigma \mathcal{T}_l)^2},  \quad l\geqslant 1, \quad \sigma=\pm 1, \quad \eta=\pm 1
\label{L1mode}
\eea

Adding and subtracting the vacuum term in (\ref{omega-init}), one can write the thermodynamic potential as the sum of vacuum ($\Omega_{vac}$), zero-temperature ($\Omega_{\mu}$), and finite-temperature ($\Omega_\beta$) contributions,
\begin{equation}\label{omega-2}
 \Omega(B,\mu,T)=\Omega_{vac}+\Omega_{\mu} +\Omega_\beta
\end{equation}
with
\begin{equation}\label{Omega-vac}
 \Omega_{vac} = \Omega(B,0,0)=- \frac{eB}{8\pi^2} \int_{-\infty}^{\infty}  dp_3  \sum_{\eta\sigma l} \vert E_{\eta\sigma l}\vert ,
\end{equation}
\begin{equation}\label{Omega-mu}
\Omega_{\mu} = \Omega(B,\mu,0)=-\frac{eB}{8\pi^2} \int_{-\infty}^{\infty}  dp_3  \sum_{\eta\sigma l}( \vert E_{\eta\sigma l}-\mu\vert-\vert E_{\eta\sigma l}\vert ),
\end{equation}
\begin{equation}\label{Omega-beta}
 \Omega_\beta=  \Omega(B,0,T)=-\frac{eB}{4\pi^2} \int_{-\infty}^{\infty}  dp_3  \sum_{\eta\sigma l} \frac{1}{\beta} \ln(1+e^{-\beta\vert E_{\eta\sigma l}  -\mu\vert})
\end{equation}

As shown in Section II, in the strong-field approximation all the $\mathcal{T}_l$ for $l>1$ are very small and can be neglected. In this approximation all the energy modes, except for $l=1$, reduce to the modes of the undressed theory.

In Appendix \ref{appendixA-1}, we give details on the calculation and carry out the renormalization of the vacuum term (\ref{Omega-vac}) in the strong-field region to obtain the following renormalized thermodynamic potential in the strong-field approximation
\bea
\Omega^{S}_R= \Omega_{vac} ^{S(R)}+\Omega_{\mu}^S +\Omega_\beta^S,
    \label{omega-3}
\eea
with
\bea
   \Omega^{S(R)}_{vac} =\frac{1}{8\pi^2}\int_{1/\Lambda^2}^{\infty} ds \frac{e^{-sm^2}}{s^3}\left[eBs\coth(eBs)-1- \frac{(eBs)^2}{3}\right] +\frac{eB}{4\pi^2} \mathcal{T}_1^2\left[\ln\frac{2eB}{m^2}+2\right],
 \label{omega-v}
\eea

\bea
   \Omega_{\mu}^S=&-&\frac{eB}{4\pi^2}\Bigg\{ \theta(\mu-m)\left[\mu\sqrt{\mu^2-m^2}-m^2 \ln\frac{\mu+\sqrt{\mu^2-m^2}}{m}\right] \nonumber
   \\
&+&\theta(\mu-M_1^+)\left[\mu\sqrt{\mu^2-{M_1^+}^2}-{M_1^+}^2 \ln\frac{\mu+\sqrt{\mu^2-{M_1^+}^2}}{M_1^+}\right]\nonumber
 \\
  &+&\theta(\mu-M_1^-)\left[\mu\sqrt{\mu^2-{M_1^-}^2}-{M_1^-}^2 \ln\frac{\mu+\sqrt{\mu^2-{M_1^-}^2}}{M_1^-}\right] \nonumber
  \\
 &+&2\sum_{l=2}^{\infty}\theta(\mu-M_l)\left[\mu\sqrt{\mu^2-{M_l}^2}-{M_l}^2 \ln\frac{\mu+\sqrt{\mu^2-{M_l}^2}}{M_l}\right] \Bigg\},
\label{omega-0}
\eea
and
\bea
   \Omega_\beta^S=&-&\frac{eB}{2\pi^2\beta} \int_{0}^{\infty}dp_3\ln(1+e^{-\beta|\sqrt{p_3^2 +m^2}+\mu|})(1+e^{-\beta|\sqrt{p_3^2 +m^2}-\mu|})
    \nonumber
   \\
&-&\frac{eB}{2\pi^2\beta}\int_{0}^{\infty}dp_3\ln(1+e^{-\beta|\sqrt{p_3^2 +{M_1^+}^2}+\mu|})(1+e^{-\beta|\sqrt{p_3^2 +{M_1^+}^2}-\mu|}) \nonumber
 \\
 &-&\frac{eB}{2\pi^2\beta} \int_{0}^{\infty}dp_3\ln(1+e^{-\beta|\sqrt{p_3^2 +{M_1^-}^2}+\mu|})(1+e^{-\beta|\sqrt{p_3^2 +{M_1^-}^2}-\mu|})\nonumber
  \\
&-&\frac{eB}{\pi^2\beta}\sum_{l=2}^{\infty}\int_{0}^{\infty}dp_3\ln(1+e^{-\beta|\sqrt{p_3^2 +{M_l}^2}+\mu|})(1+e^{-\beta|\sqrt{p_3^2 +{M_l}^2}-\mu|}),
 \label{omega-beta}
\eea
where
\bea
M_1^\pm=\sqrt{2eB+m^2}\pm \mathcal{T}_1, \qquad M_l=\sqrt{2eBl+m^2}
\label{effect-mass1}
\eea

The leading vacuum contribution to $\Omega$ in the strong-field approximation (see Appendix \ref{appendixA-1} for details) is then
\bea\label{vac-Ren-App}
\Omega_{vac}^{S(R)}&=&-\frac{\alpha B^2}{6\pi}\ln\left(\frac{eB}{m^2}\right)+\frac{eB}{4\pi^2} \mathcal{T}_1^2\left[\ln\frac{2eB}{m^2}+2\right]
\nonumber
 \\
 &=&-\frac{\alpha B^2}{6\pi}\ln\left(\frac{eB}{m^2}\right)+\frac{eB}{4\pi^2} \left[\left(\frac{\alpha m}{8\pi}\right)\ln^2{\left(\frac{2eB}{m^2}\right)}\right]^2\left[\ln\frac{2eB}{m^2}+2\right],
\eea
where  in the second line $\mathcal{T}_1$ was evaluated using (\ref{T1}), and the first term was obtained in \cite{Ragazzon} calculating the effective potential at strong field and neglecting the AMM.

Here the following comment is in order. The contribution of the AMM to the vacuum part of the thermodynamic potential has been previously calculated in \cite{DittrichAMM} with the help of the Green's function. In the strong-field region the result reported in \cite{DittrichAMM}  was
\begin{equation}\label{Dittrich}
\hat{\Omega}_{vac}^{S(R)}=-\frac{\alpha B^2}{6\pi}\ln\left(\frac{eB}{m^2}\right)-\frac{(eB)^2}{32\pi^2}\left(\frac{\alpha}{2\pi}\right)^2\ln\left(\frac{eB}{m^2}\right),
\end{equation}
Clearly, there is a discrepancy between (\ref{Dittrich}) and (\ref{vac-Ren-App}), the origin of which can be traced back to some inconsistencies in the treatment followed in  \cite{DittrichAMM} to obtain the strong-field result. On the one hand, Ref. \cite{DittrichAMM} considered the AMM found by Schwinger and used it in calculations at arbitrary field strength, including the strong field region, despite that as already discussed, in this region the AMM becomes a very different function of the magnetic field and depends on the LL. On the other hand, as described in Appendix B, several steps of the calculations done in \cite{DittrichAMM} were only valid for weak fields, however, they were used indistinctly for weak and strong fields. The calculations of Ref. \cite{DittrichAMM} are then reliable in the weak-field region, but fail to describe the AMM-dependent terms in the strong-field case.

\section{AMM in the weak-field approximation}\label{section4}

To get the AMM in the weak-field limit, it will be convenient,
for the sake of clarity and to shade light in our discussion, to use
an alternative method that combines the Ritus's approach, where the LL contributions are explicit, and
the proper time formalism. The result has to coincide with that found
by Schwinger \cite{Proper-Time} by using an independent method. In
doing this, we will stress the steps where the weak-field
approximation becomes a basic element of the derivation. Also, it will be apparent the different behavior of the AMM's for the different LL's in each approximation.

Due to the fact that the energy separation between consecutive LL's
is proportional to $\sqrt{2eB}$, it is expected that at weak field all
Landau levels will contribute on equal footing to the fermion
self-energy. Then, we will find convenient, in order to take into
account the contributions of all the LL's into the fermion
self-energy, to work with the self-energy operator in the
configuration space (\ref{SE-coordinate}) with the field dependent
fermion Green's function given in the Ritus's approach as
\bea
    G(x,y)=\sumint\frac{d^4p}{(2\pi)^4}
           \mathbb{E}_{p}(x)
          {G}^{l}(\overline{p})
           \overline{\mathbb{E}}_{p}(y).
\label{fulllan2}
\eea
where
\bea
 {G}^{l}(\overline{p})=\frac{1}{\slsh{\overline{p}}-m}\Pi(l)
\eea
is the fermion propagator in momentum space and we introduced the notation
\bea
  \sumint\frac{d^4p}{(2\pi)^4}\equiv\sum_{l=0}^\infty\int\frac{dp_0dp_2dp_3}{(2\pi)^4}
\eea

We can rewrite the electron propagator in~(\ref{fulllan2}) as
\bea
    G(x,y)
     &=&(\slsh{\Pi_x}+m)\sumint\frac{d^4p}{(2\pi)^4}
           \frac{\mathbb{E}_{p}(x)\Pi(l)\overline{\mathbb{E}}_{p}(y)}
                {\overline{p}^2-m^2}
\label{fulllan3b}
\eea
where ${\Pi_x}^\mu=i{\partial_x}^\mu-eA^\mu$
and we used the property,
$\slsh{\Pi_x}\mathbb{E}_p(x)=\mathbb{E}_p(x)
\slsh{\overline{p}}$, satisfied by the Ritus eigenfunctions, $ \mathbb{E}_{p}(x)$, defined in (\ref{Ep})-(\ref{E-x}).

Now, to perform the summation over all Landau levels, we use the
proper-time representation and the integral representation for the
parabolic cylinder functions, so, we get (See
Appendix~\ref{appendixB} for details)
\bea
 G(x,y)&=&-(\slsh{\Pi_x}+m)\frac{\Phi(x,y)}{(4\pi)^2}
    \int_0^\infty \frac{eBds}{s \sin(eBs)}\ e^{-is(m^2-i\epsilon)}
          e^{-i[\frac{1}{4s}(x-y)_{||}^2-\frac{eB}{4\tan(eBs)}(x-y)_\perp^2]}
\nonumber \\
    &&\times\left[e^{ieBs}\Delta(+)+e^{-ieBs}\Delta(-)
    \right]
\label{summedprop2}
\eea
where
$\Phi(x,y)=\exp\left[{i\frac{eB}{2}(x_2-y_2)(x_1+y_1)}\right]$
is the well known Schwinger's phase (recall that $F_{21}=B$)  \cite{Chodos}.

In the weak-field approximation $(eB\ll m^2)$, we can perform a Taylor expansion
up to linear terms in $eB$  in the integrand of \Eq{summedprop2}, to find
\bea
 G(x,y)
&\simeq&-\frac{\Phi(x,y)}{(4\pi)^2}
    \int_0^\infty \frac{ds}{s^2}\ e^{-is(m^2-i\epsilon)}
    e^{-i\frac{(x-y)^2}{4s}}
     \left(
     \frac{\slsh{x}-\slsh{y}}{2s}
      +\frac{e}{2}\gamma^\mu F_{\mu\nu}(x-y)^\nu+m
     \right)
\nonumber \\
    &&\times\left[1+ieBs(\Delta(+)-\Delta(-))
    \right],
\label{weakprop1}
\eea
where we used the fact that the Schwinger phase satisfies the identity
\bea
   {\Pi^x}_\mu\Phi(x,y)&=&\frac{e}{2}F_{\mu\nu}(x-y)^\nu \Phi(x,y).
\label{phaseeq1}
\eea

With the help of the identities
$eB(\Delta(+)-\Delta(-))=-\frac{e}{2}\sigma^{\mu\nu}F_{\mu\nu}$
and
$
   \left[\gamma_\rho,\sigma_{\mu\nu}\right]F^{\mu\nu}X^\rho
    =-4i\gamma^\mu F_{\mu\rho}X^\rho
$ with $X^\rho$ an arbitrary four-vector, we rewrite \Eq{weakprop1}
as
\bea
 G(x,y)
&=&-\frac{\Phi(x,y)}{(4\pi)^2}
    \int_0^\infty \frac{ds}{s^2}\ e^{-is(m^2-i\epsilon)}
    e^{-i\frac{(x-y)^2}{4s}}
\nonumber \\
&&\times
     \left(
     \frac{\slsh{x}-\slsh{y}}{2s}
      +\frac{1}{4}\left[\slsh{x}-\slsh{y}\ ,\ i\frac{e}{2}\sigma^{\mu\nu} F_{\mu\nu}\right]+m
     \right)
\left[1-si\frac{e}{2}\sigma^{\mu\nu}F_{\mu\nu}
    \right]
\nonumber \\
&\simeq&-\frac{\Phi(x,y)}{(4\pi)^2}
    \int_0^\infty \frac{ds}{s^2}\ e^{-is(m^2-i\epsilon)}
    e^{-i\frac{(x-y)^2}{4s}}
\nonumber \\
&&\times \left(
     \frac{\slsh{x}-\slsh{y}}{2s}
      -\frac{1}{4}\left\{\slsh{x}-\slsh{y}\ ,\ i\frac{e}{2}\sigma^{\mu\nu} F_{\mu\nu}\right\}+m
-msi\frac{e}{2}\sigma^{\mu\nu}F_{\mu\nu}
    \right)
\label{weakprop1a}
\eea
where in the last line we kept up to linear terms in $eB$, in agreement with the weak-field approximation.

To find the fermion self-energy we have to substitute in~(\ref{SE-coordinate}) the photon propagator
in configuration space
\bea
   D_{\mu\nu}(x-y)&=& \frac{-ig_{\mu\nu}}{(4\pi)^2}
    \int_0^\infty \frac{dt}{t^2}\
    e^{-i\frac{(x-y)^2}{4t}}       ,
\eea
together with the fermion propagator~(\ref{weakprop1a}). Then, after using the identities
$\gamma^\mu\gamma^\rho\gamma_\mu=-2\gamma^\rho$, \
$\gamma^\mu\gamma^\alpha\gamma^\beta\gamma_\mu=4g^{\alpha\beta}$ and
$\gamma^\mu\gamma^{\rho}\gamma^\alpha\gamma^\beta\gamma_\mu=-2\gamma^{\beta}\gamma^\alpha\gamma^\rho$,
the fermion self-energy reduces to
\bea
   \Sigma(x,y)
   &=&i\frac{e^2\Phi(x,y) }{(4\pi)^4}
    \int_0^\infty \frac{dsdt}{s^2t^2}\ e^{-is(m^2-i\epsilon)}
    e^{-i(\frac{1}{s}+\frac{1}{t})\frac{(x-y)^2}{4}}
\nonumber \\
    &&\times
\left(-\frac{\slsh{x}-\slsh{y}}{s}
     -\frac{1}{2}\left\{\slsh{x}-\slsh{y}\ ,\ i\frac{e}{2}\sigma^{\mu\nu} F_{\mu\nu}\right\}
     +4m
     \right)
\label{anomalous3}
\eea
Now, we introduce a new variable
  defined as
\bea
   \frac{1}{w}=\frac{1}{s}+\frac{1}{t}
\eea
to rewrite \Eq{anomalous3} as follows
\bea
   \Sigma(x,y)
   &=&i\frac{e^2\Phi(x,y)}{(4\pi)^4}
    \int_0^\infty\frac{ds}{s^2}
    \int_0^s\frac{dw}{w^2}\ e^{-is(m^2-i\epsilon)}
    e^{-i\frac{(x-y)^2}{4w}}
\nonumber \\
    &&\times
 \left(-\frac{\slsh{x}-\slsh{y}}{s}
     -\frac{1}{2}\left\{\slsh{x}-\slsh{y}\ ,\ i\frac{e}{2}\sigma^{\mu\nu} F_{\mu\nu}\right\}
     +4m
     \right)
\label{anomalous4}
\eea

On the other hand, from \Eq{phaseeq1} we have
\bea
    \left(\slsh{\Pi^x}
-\frac{1}{4}\left[\slsh{x}-\slsh{y}\ ,\ i\frac{e}{2}\sigma^{\mu\nu}
      F_{\mu\nu}\right]
\right)
 \Phi(x,y)e^{-i\frac{(x-y)^2}{4w}}
    &=&\Phi(x,y)e^{-i\frac{(x-y)^2}{4w}}\left[\frac{\slsh{x}-\slsh{y}}{2w}\right],
\label{anomalous5}
\eea

Thus, solving this self-consistent equation
iteratively in the weak-field limit, the leading terms have
the form
\bea
    \Phi(x,y)e^{-i\frac{(x-y)^2}{4w}}\left[\slsh{x}-\slsh{y}\right]
    &\simeq&
2w\left(\slsh{\Pi}^x
       -\frac{1}{4}
        \left[2w\slsh{\Pi}^x\ ,\ i\frac{e}{2}\sigma^{\mu\nu}F_{\mu\nu}\right]
  \right)
    \Phi(x,y)e^{-i\frac{(x-y)^2}{4w}}.
\label{anomalous6}
\eea

Replacing \Eq{anomalous6} into  \Eq{anomalous4}, we obtain
\bea
   \Sigma(x,y)
   &\simeq&i\frac{e^2}{(4\pi)^2}
    \int_0^\infty\frac{ds}{s^2}
    \int_0^s dw\ e^{-is(m^2-i\epsilon)}
     \left(-\frac{2w}{s}
\left(\slsh{\Pi}^x
       -\frac{w}{2}
        \left[\slsh{\Pi}^x\ ,\ i\frac{e}{2}\sigma^{\mu\nu}F_{\mu\nu}\right]
         \right)
     \right.
\nonumber \\
   &&\left.
-\frac{1}{2}\left\{2w \slsh{\Pi}^x,\ i\frac{e}{2}\sigma^{\mu\nu} F_{\mu\nu}\right\}
     +4m\right)e^{iw\frac{e}{2}\sigma_{\mu\nu}F^{\mu\nu}}
     <x|U|y>,
\label{anomalous8}
\eea
where we used that the space-representation of the proper-time
evolution operator  $U=e^{i\mathcal{H}w}$, with $\mathcal{H}={(\Pi_\mu)}^2$, is given by
\bea
 <x|U|y>\equiv\frac{1}{ (4\pi)^2w^2}\Phi(x,y)e^{-i\frac{(x-y)^2}{4w}}e^{-iw\frac{e}{2}\sigma_{\mu\nu}F^{\mu\nu}}.
\eea

Multiplying Eq. (\ref{anomalous8}) by $\psi(y)$, a solution of the
Dirac equation, $[\slsh{\Pi}-m]\psi=0$, and integrating over $y$, we
arrive at
\bea
  \int d^4y\Sigma(x,y)\psi(y)
       &=&i\frac{e^2}{(4\pi)^2}
    \int_0^\infty\frac{ds}{s^2}
    \int_0^s dw\ e^{-is(m^2-i\epsilon)}
     \left(-\frac{2w}{s}
\left(\slsh{\Pi}^x
       -\frac{w}{2}
        \left[\slsh{\Pi}^x\ ,\ i\frac{e}{2}\sigma^{\mu\nu}F_{\mu\nu}\right]
 \right)
     \right.
\nonumber \\
   &&\left.
-w\left\{\slsh{\Pi}^x,\ i\frac{e}{2}\sigma^{\mu\nu} F_{\mu\nu}\right\}
     +4m\right)e^{iw\frac{e}{2}\sigma_{\mu\nu}F^{\mu\nu}}
     \psi(x)e^{iwm^2}
\label{anomalous9}
\eea
where we used the fact that
\bea
   \int d^4y\ <x|U|y>\psi(y)=\int d^4y\ <x|U|y><y|\psi>=<x|U|\psi>=<x|\psi>e^{iwm^2}
\label{anomalous10}
\eea

Bearing in mind the identity  $[A,B]\equiv \{A,B\}-2BA$, then \Eq{anomalous9} becomes
\bea
  \int d^4y\Sigma(x,y)\psi(y)
       &=&i\frac{e^2}{(4\pi)^2}
    \int_0^\infty \frac{dsdw}{s^2}\ e^{-is(m^2-i\epsilon)}
    \left(4m
-\frac{2w}{s}\slsh{\Pi^x}
+4mwi\frac{e}{2}\sigma_{\mu\nu}F^{\mu\nu}
 \right.
\nonumber \\
   &&-\left.
      w\left(1+\frac{w}{s}\right)
      \left\{
         \slsh{\Pi}^x,
     i\frac{e}{2}\sigma_{\mu\nu}F^{\mu\nu}\right\}
     \right)
     \psi(x)e^{iwm^2}
\label{anomalous9b}
\eea

Next, using once again that $\slsh{\Pi}\psi=m\psi$ in (\ref{anomalous9b}),
we have
\bea
  \int d^4y\Sigma(x,y)\psi(y)
 &=&i\frac{e^2}{(4\pi)^2}
    \int_0^\infty \frac{dsdw}{s^2}\ e^{-is(m^2-i\epsilon)}
\nonumber \\
   &&\times
    \left[2m\left(2-\frac{w}{s}\right)
     + 2mw\left(1-\frac{w}{s}\right)
     i\frac{e}{2}\sigma_{\mu\nu}F^{\mu\nu}
     \right]
     \psi(x)e^{iwm^2}
\label{finaleq}
\eea
where we kept only the leading contribution in the magnetic field
$\mathcal{O}(eB)$.

Eq. (\ref{finaleq}) can be written as
\begin{equation}
\Sigma(x)\psi(x)=(mI-\mathcal{T}i\gamma^1\gamma^2)\psi(x)
\end{equation}\label{struct}
Here, the two terms corresponding to the two independent Dirac
structures, are the radiative mass and AMM, respectively given by
\bea
  m&\equiv&m_0+\Sigma \\
    &=&m_0+\frac{\alpha}{2\pi}m
        \int_0^\infty\frac{ds}{s^2}
    \int_0^s dw\ e^{-i(s-w)(m^2-i\epsilon)}
     \left(2-\frac{w}{s}
     \right)
\label{massterm} \\
  \mathcal{T}
     &\equiv&\frac{\alpha}{2\pi}
 \int_0^\infty \frac{dsdw}{s}\ e^{-i(s-w)(m^2-i\epsilon)}
     \frac{w}{s}\left(1-\frac{w}{s}\right)eBm
\label{anomalosandmass}
\eea
with $m_0$ the electron bare mass.
These two equations coincide with Schwinger's results for the mass
and AMM Ref.~\cite{Proper-Time}.

Following Ref.~\cite{Proper-Time}, the last two equations can be
evaluated by using a new variable $u\equiv 1-w/s$ and  making the
replacement $s\rightarrow -is $, so, we arrive at
\bea
   m&=& m_0+\frac{3}{4}\frac{\alpha}{\pi}m
        \left(\int_0^\infty\frac{ds}{s}e^{-sm^2}+\frac{5}{6}\right)\\
  \mathcal{T}
     &=&\frac{\alpha}{2\pi}\frac{eB}{2m}
\label{anomalosandmass2}
\eea
where $m$ is the electron's renormalized mass.

As we can explicitly see in the derivations of this section, the formula (\ref{anomalosandmass2}) for the AMM is only valid in the weak-field limit ($eB<m^2$).

\section{Thermodynamic Potential with AMM in the Weak-Field
  Limit}\label{section5}

In this Section we aim to find the thermodynamic potential of the effective theory in a weak magnetic field. Since the field is weak, all the LLs contribute to the AMM, so the AMM entering in the effective theory is now the one found by Schwinger many years ago and given by the equation (\ref{anomalosandmass2}). It is well known that the Euler-Heisenberg Lagrangian (i.e. the vacuum contribution to the thermodynamic potential) for charged fermions in the weak-field limit contains divergences associated with the renormalization of the field and charge \cite{Proper-Time}. A peculiarity of the present case is the appearance of new divergencies connected to the AMM.

Following the same approach of Section III, we can write the renormalized thermodynamic potential in the weak-field limit as the sum of the vacuum ($\Omega_{vac} ^{W(R)}$), zero-temperature ($\Omega_{\mu}^W $) and finite-temperature ($\Omega_\beta^W$) contributions
\bea
\Omega^{W}_R= \Omega_{vac} ^{W(R)}+\Omega_{\mu}^W +\Omega_\beta^W,
    \label{omega-3-weak}
\eea
where
\bea
   \Omega_{vac}^{W(R)}&=&
   -\frac{1}{2(2\pi)^2}
   \int_0^\infty
   \frac{ds}{s^3}
   e^{-s\left(m^2+\mathcal{T}^2\right)}
   {\Bigg\{ }
   -|eB|s\sinh(2s\mathcal{T}m)
\nonumber \\
&+&
   \frac{\sqrt{\pi}}{2\pi i}
        \int_{\gamma-i\infty}^{\gamma+i\infty}
           \frac{dt}{t^{\frac{1}{2}}}e^{t+\frac{s^2\mathcal{T}^2m^2}{t}}
    |eB|s \coth\left(s\left(1-\frac{s\mathcal{T}^2}{t}\right)eB\right)
\nonumber \\
   &-&      \left(1+s\mathcal{T}^2+2s^2 m^2\mathcal{T}^2
       -2ms^2\mathcal{T}eB- \frac{1}{6}s^2\mathcal{T}^4+\frac{(eBs)^2}{3}
     \right)
   \Bigg\},
\label{poteff10new}
\eea
\bea
   \Omega^{W}_\mu=-\frac{|eB|}{(2\pi)^2}
   \int dp_{3}
 \left\{
   (\mu-\varepsilon_{++0})\theta(\mu-\varepsilon_{++0})
   +
   \sum_{\sigma=\pm1}
   \sum_{l=1}^{\infty}
          (\mu-\varepsilon_{+\sigma l})\theta(\mu-\varepsilon_{+\sigma l})
   \right\},
\eea
and
\bea
   \Omega^W_{\beta}
   &=&-\frac{|eB|}{(2\pi)^2\beta}
   \int dp_{3}
      \ln\left(1+e^{-\beta |\varepsilon_{++0}+\mu|}\right)
         \left(1+e^{-\beta |\varepsilon_{++0}-\mu|}\right)
\nonumber\\
   &&-\frac{|eB|}{(2\pi)^2\beta}\sum_{\sigma=\pm1}
   \sum_{l=1}^{\infty}
   \int dp_{3}
      \ln\left(1+e^{-\beta |\varepsilon_{+\sigma l}+\mu|}\right)
         \left(1+e^{-\beta |\varepsilon_{+\sigma l}-\mu|}\right),
\label{poteffdens2new}
\eea
with
\bea
   \varepsilon_{\eta\sigma l}
   \equiv  \eta\sqrt{p_{3}^2+\left(\sqrt{m^2+2leB}+\sigma\mathcal{T}\right)^2}
\label{varepsilondef1new}
\eea
and $\mathcal{T}$ given in (\ref{anomalosandmass2}). The details of these calculations are given in Appendix D.

Note that in the above equations the AMM  does not contain any
dependence on the LL's. This is one of the main differences with respect to the
strong-field case.

In the weak-field expansion, the leading terms of the renormalized
potential (\ref{poteff10new}), taken  up to $\mathcal{O}((eB)^4)$
order and $\alpha^2$ correction, are (see Appendix~D)
\bea
   \Omega_{vac}^{W(R)}(B,0,0)=\frac{1}{2(2\pi)^2}
    \left\{\frac{(eB)^4}{45m^4}
         +\frac{(eB)^2\mathcal{T}^2}{3m^2} \right\}
\label{poteffRen-2}
\eea
Here, we should mention that if in the effective potential found in
Ref. \cite{DittrichAMM} we take the weak-field limit up to the same
order we are considering here, we obtain the same result,
although in the calculation carried out in Ref. \cite{DittrichAMM}
there are certain inconsistencies as we discuss in  Appendix
B. This is indicating that the inconsistencies in the
calculations of \cite{DittrichAMM} did not affect the results in the
weak-field limit, but only those at strong field.

Similarly, the leading terms in the weak-field approximation up to $\mathcal{O}((eB)^2)$
order and $\alpha^2$ correction of the finite-density, zero-temperature thermodynamic potential reads
\bea\label{TP-munew}
  \Omega^W_\mu
    &=&-\frac{1}{8\pi^2}
 \Bigg\{\frac{2}{3}\mu p_F^3-m^2\mu p_F
       +m^4\ln\left(\frac{p_F+\mu}{m}\right)\Bigg\}\theta(\mu-m)
\nonumber \\
      &&-\frac{\mathcal{T}^2}{4\pi^2}\left(\mu p_F+m^2\ln\left(\frac{p_F+\mu}{m}\right) \right)\theta(\mu-m)
   +\frac{|eB|m\mathcal{T}}{2\pi^2}\ln\left(\frac{p_F+\mu}{m}\right)\theta(\mu-m)
\nonumber \\
   &&
    -\frac{|eB|^2}{12\pi^2}\ln\left(\frac{p_F+\mu}{m}\right)\theta(\mu-m)
\eea
where $p_F\equiv\sqrt{\mu^2-m^2}$. In the above equation
the first three terms are the contribution to the thermodynamic
potential at $B=T=0$ of a system of fermions at finite
density~\cite{Kapusta}. The next three terms are the AMM contributions
to the magnetized fermion thermodynamic potential at finite density and zero temperature. Finally, the last term accounts for the pure magnetic
contribution. Comparing (\ref{poteffRen-2}) and (\ref{TP-munew}), we can see that at weak field ($eB\ll \mu^2$), the vacuum contribution can be neglected with respect to the finite-density contribution at zero temperature.

\section{Equation of State with AMM}\label{section7}

In this section we will investigate the effect of the AMM in the EoS. We will separate the analysis for the weak-field and strong-field regimes using the expressions of the thermodynamic potential at $T=0$ found in each limit, as well as the corresponding formula of the AMM. Also, to show the consequences of using the wrong prescription of taking the AMM in the Schwinger approximation when one is dealing with strong-field calculations, we will find the deviation in the parallel pressure produced by using $\mathcal{T}$ in place of $\mathcal{T}_1$.

\begin{figure}[!ht]
\begin{center}
\includegraphics[width=12cm,height=10cm]{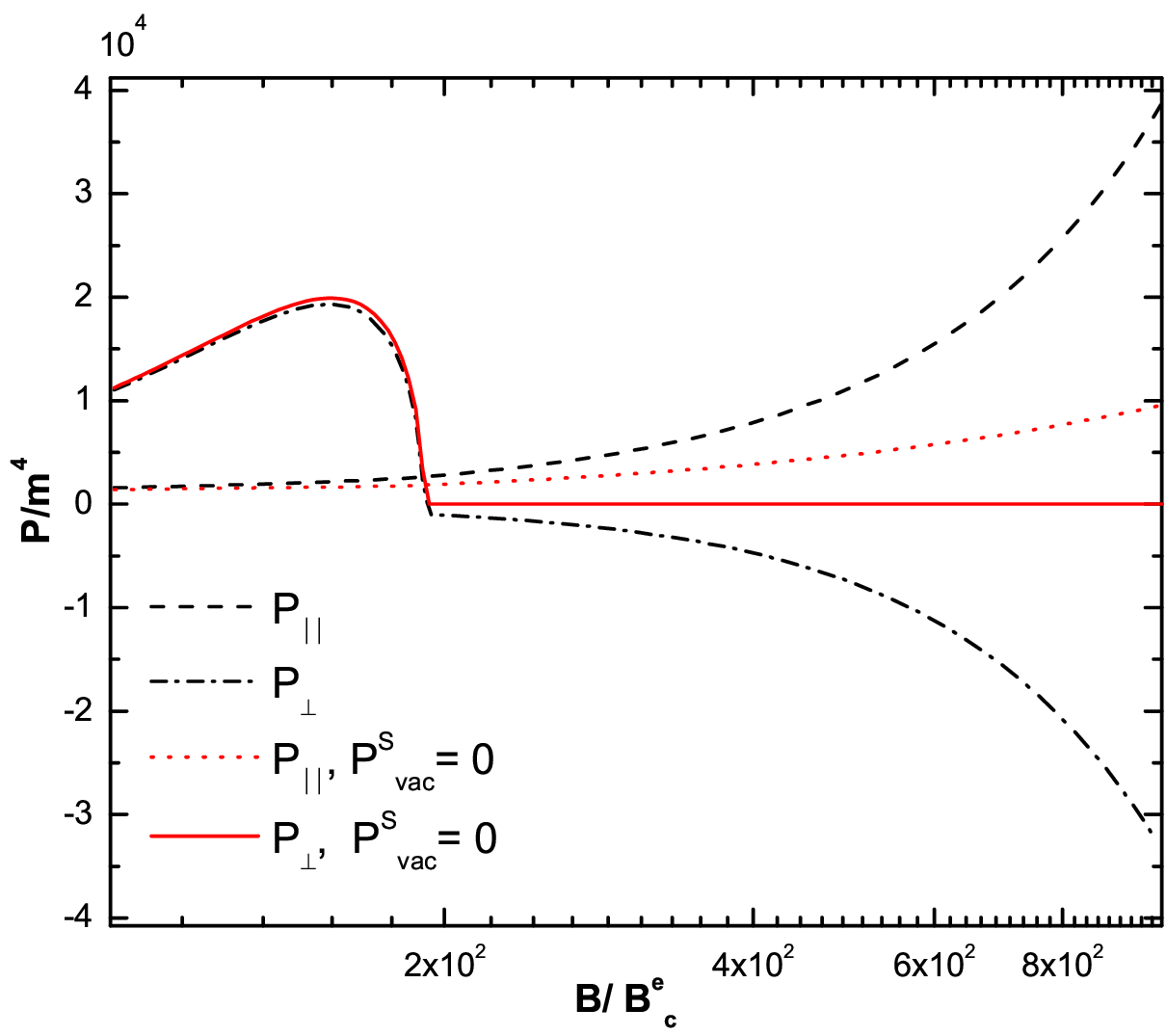}
\caption{\footnotesize (Color online) \emph{Vacuum effects on the parallel and transverse pressures}. Profile of the parallel and transverse pressures versus $B/B_c^e$ in the
 strong-field region, for $\mu=10.0$~MeV and $m/\mu\simeq0.051$.  All the plots include the AMM, but ignore the Maxwell pressure. Plots with vacuum contribution included (black) and ignored (red) are compared. The figure shows that the vacuum terms in the two pressures become progressively relevant with increasing magnetic field.  In the region of very strong fields the vacuum terms are big enough to produce a sizable negative contribution to the perpendicular pressure.} \label{fig3}
\end{center}
\end{figure}

As  pointed out in the Introduction, the breaking of the rotational symmetry produced by a uniform magnetic field gives rise to an anisotropy in the energy-momentum tensor \cite{Landau-Lifshitz} that leads to a splitting of the pressure into two different components  \cite{Canuto}-\cite{EOS-Ferrer}, one along the field (longitudinal pressure, $P_\parallel$) and another in the perpendicular direction (transverse pressure, $P_\bot $). Those pressures and the energy density are respectively given by
\begin{equation}\label{E0S-Eqs}
P_\parallel = -\Omega_f^R-B^2/2, \quad  P_\bot = -\Omega_f^R-B\mathcal{M}+B^2/2, \quad \epsilon= \Omega^R_f+ \mu \mathcal{N}+B^2/2
\end{equation}
Here, $\Omega_f^R$ represents the fermion renormalized thermodynamic potential, given below in the strong-field, $\Omega_{S}^R$, and weak-field, $\Omega_{W}^R$, approximations; $\mathcal{M}=-(\partial \Omega_f^R/\partial B)$ is the system magnetization; $ \mathcal{N}=-(\partial \Omega^R_f/\partial \mu)$ its particle number density; and the terms proportional to $B^2/2$ are the renormalized Maxwell contributions  \cite{EOS-Ferrer, PRC82}. In the following considerations, we consider the zero-temperature limit, which is of particular interest for astrophysical applications.

\begin{figure}[!ht]
\begin{center}
\includegraphics[width=16cm,height=10cm]{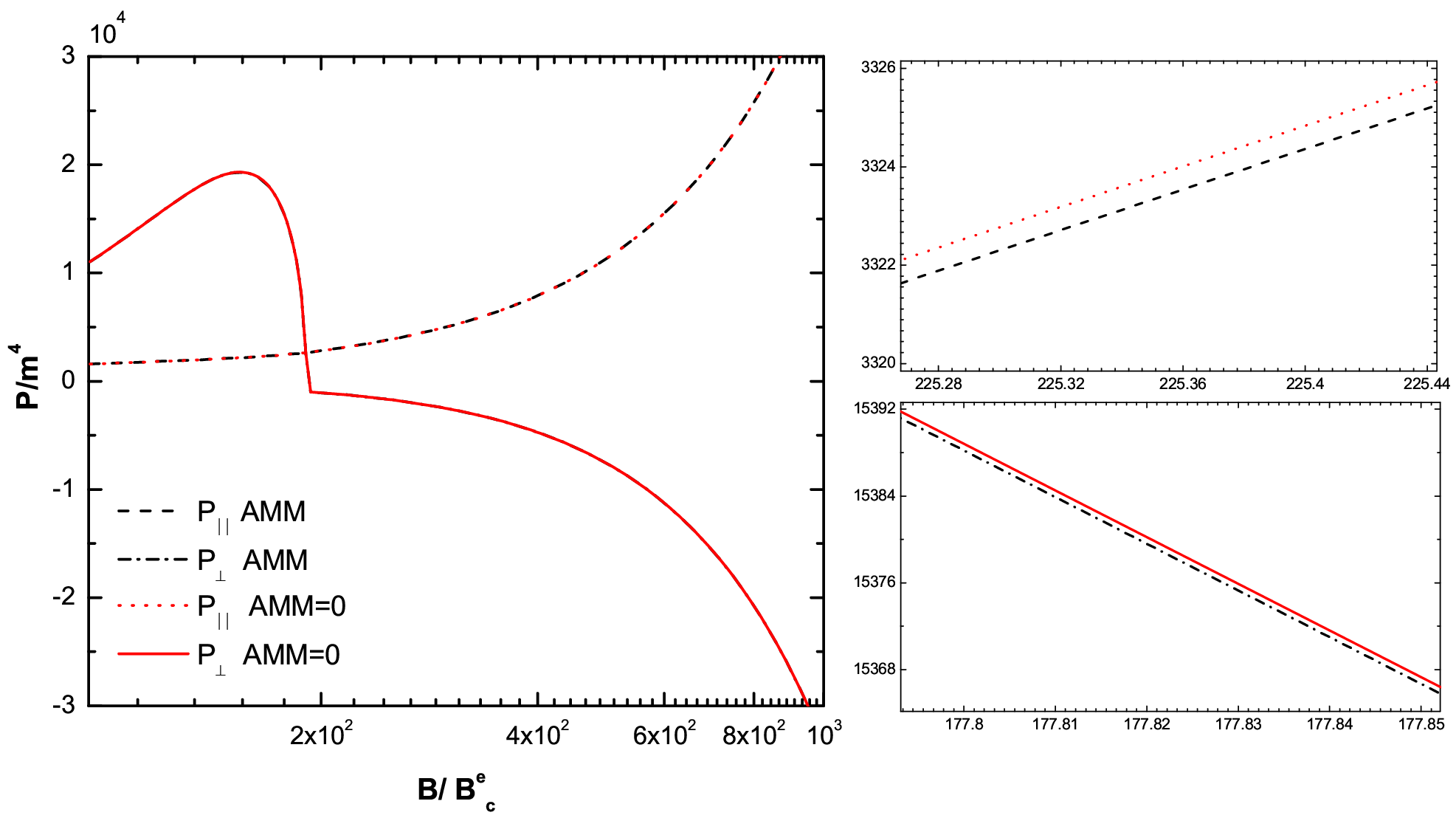}
\caption{\footnotesize (Color online) \emph{AMM effects on the parallel and transverse pressures}. Profile of the parallel and transverse pressures with the field in the strong-field region including vacuum contribution, but ignoring the Maxwell pressure. The values $\mu=10.0$~MeV and $m/\mu\simeq0.051$ were used. Black (red) curve indicates $AMM\neq0$ ($AMM=0$). The zooms in the right panels show that the AMM makes a negligibly small contribution to each pressure in the strong-field regime.} \label{fig4}
\end{center}
\end{figure}

\subsection{EoS in the strong-field approximation}

The zero-temperature limit of the thermodynamic potential in the strong-field limit, $\Omega_{S}^R$, is determined by the first two terms of (\ref{omega-3}), which are given by Eqs. (\ref{vac-Ren-App}) and (\ref{omega-0}), respectively
\bea\label{T-0}
\Omega^{S(R)}(T=0)&=&\Omega^{S(R)}_{vac}+\Omega^S_\mu=-\frac{(eB)^2}{24\pi^2}\ln\left(\frac{eB}{m^2}\right)+\frac{eB}{\pi^2}\mathcal{T}_1^2-\nonumber
\\
 &-&\frac{eB}{4\pi^2}\Bigg\{ \theta(\mu-m)\left[\mu\sqrt{\mu^2-m^2}-m^2 \ln\frac{\mu+\sqrt{\mu^2-m^2}}{m}\right] \nonumber
   \\
&+&\theta(\mu-M_1^+)\left[\mu\sqrt{\mu^2-{M_1^+}^2}-{M_1^+}^2 \ln\frac{\mu+\sqrt{\mu^2-{M_1^+}^2}}{M_1^+}\right]\nonumber
 \\
  &+&\theta(\mu-M_1^-)\left[\mu\sqrt{\mu^2-{M_1^-}^2}-{M_1^-}^2 \ln\frac{\mu+\sqrt{\mu^2-{M_1^-}^2}}{M_1^-}\right] \Bigg\}
\eea

\begin{figure}[!ht]
\begin{center}
\includegraphics[width=12cm,height=10cm]{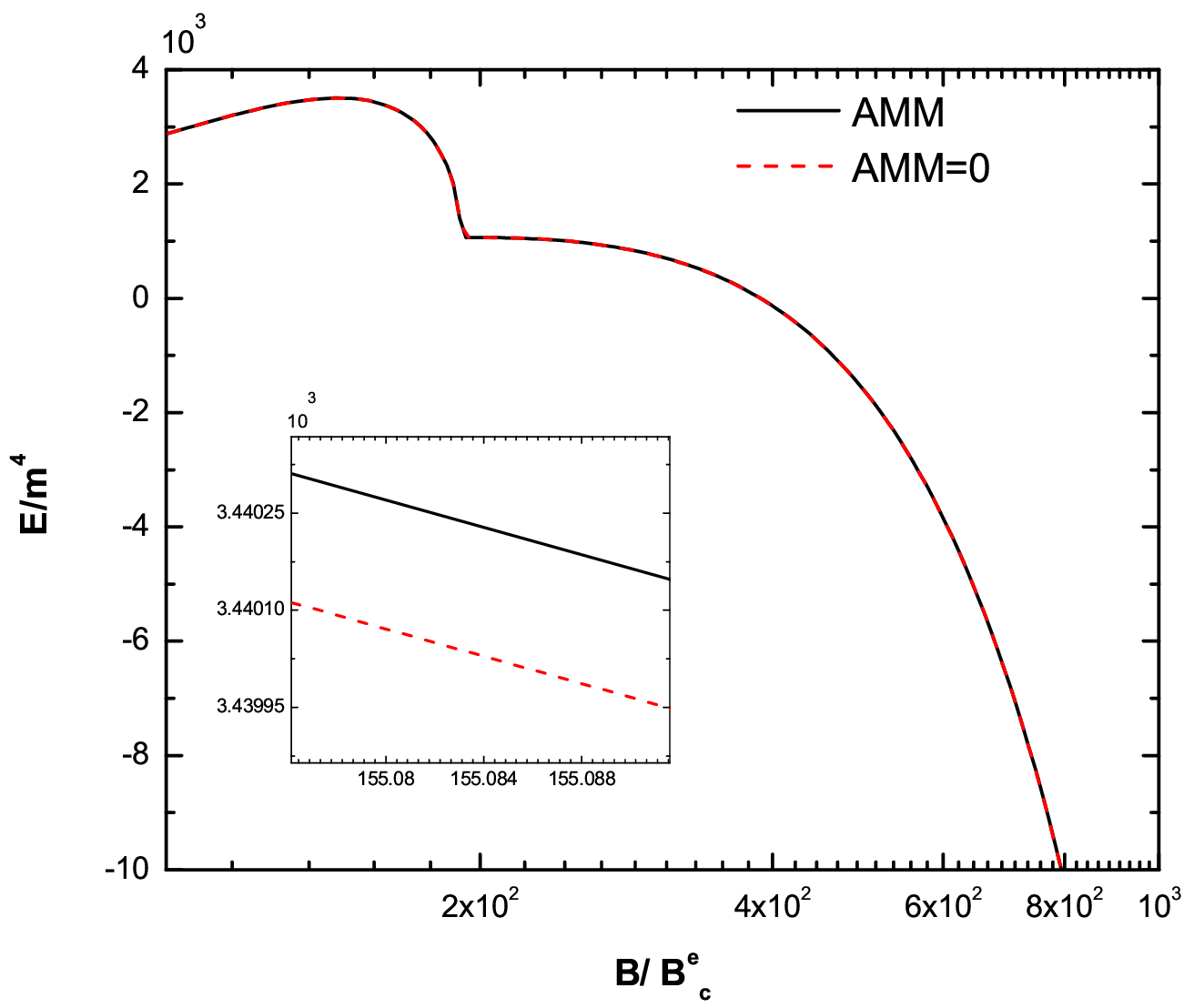}
\caption{\footnotesize (Color online) \emph{AMM effects on the energy density.} Energy density in the strong-field region including the vacuum contribution and ignoring the Maxwell one, for $\mu=10.0$~MeV and $m/\mu\simeq0.051$. Black curves indicate $AMM\neq0$, red curves indicate $AMM=0$. The zoom shows that the AMM makes no significant contribution to the energy density.} \label{fig5}
\end{center}
\end{figure}

To ensure that only the zero and first LL's are populated for the considered magnetic-field values, we should satisfy the condition $m< \mu<\sqrt{4eB+m^2}$. Using the thermodynamic potential (\ref{T-0}), we find from (\ref{E0S-Eqs}) the parallel and transverse pressures in the strong-field approximation. Their graphical representations as a function of the magnetic field are given in Fig.\ref{fig3}, without the Maxwell term. There, we compare the cases where the vacuum term is included and ignored. As can be seen, the vacuum term in the strong-field regime produces a remarkable contribution leading to a quicker turn of the transverse pressure toward negative values.

\begin{figure}[!ht]
\begin{center}
\includegraphics[width=12cm,height=10cm]{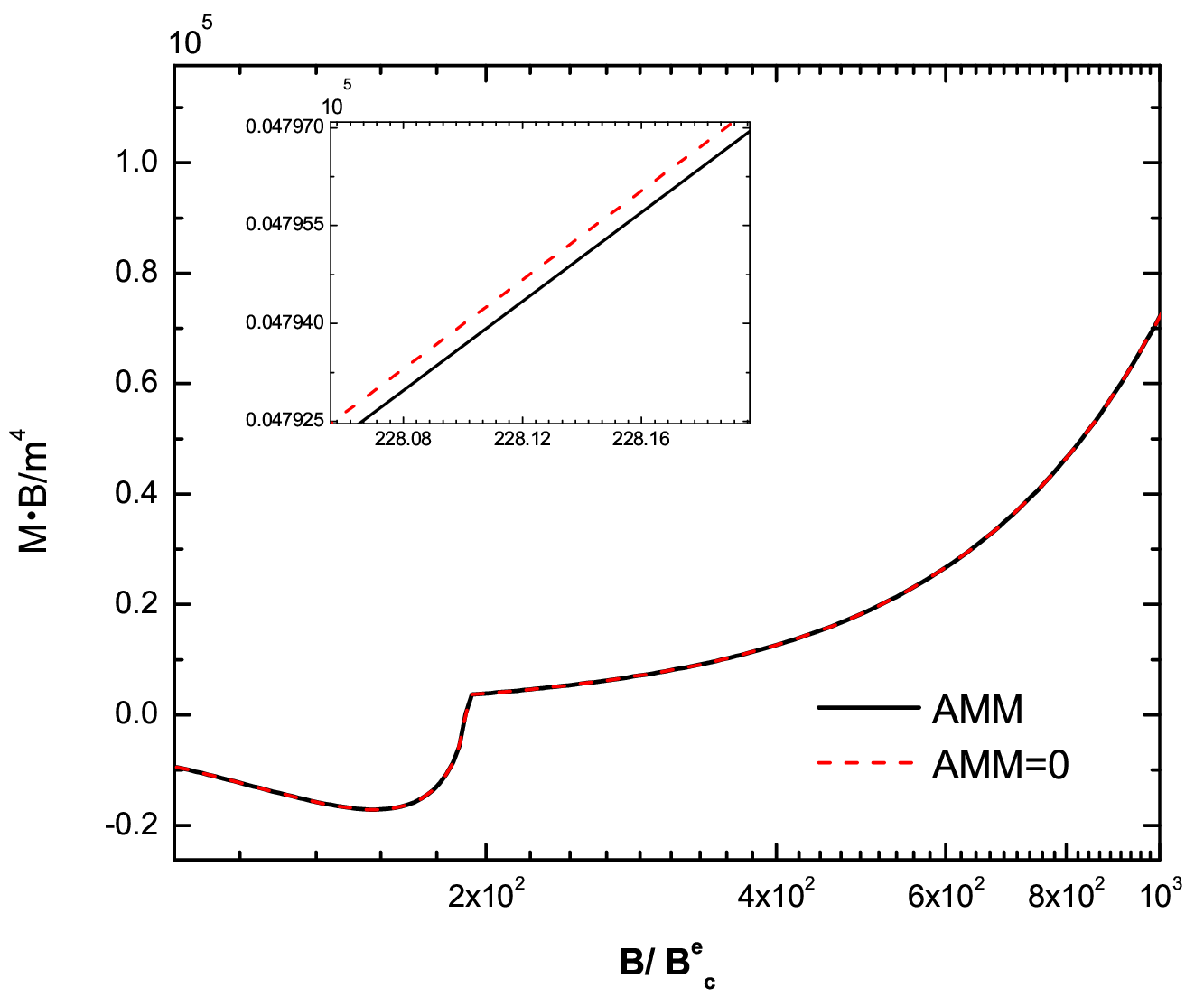}
\caption{\footnotesize (Color online) \emph{AMM effects on the magnetization.} Magnetization versus magnetic field in the strong-field region including the vacuum contribution and ignoring the Maxwell one, for $\mu=10.0$~MeV and $m/\mu\simeq0.051$.  Black curves indicate $AMM\neq0$, red curves indicate $AMM=0$. The zoom shows that the AMM makes no significant contribution to the magnetization.} \label{fig6}
\end{center}
\end{figure}

In Fig. \ref{fig4} we compare the parallel and transverse pressures, with and without the inclusion of the AMM. In both cases the vacuum and medium contributions are included, but the Maxwell term is ignored. As can be seen in the zooms of the right panel, the AMM in the strong-field regime makes a negligibly small contribution to the pressures, since it enters as an $\alpha$-correction in the thermodynamic potential.

\begin{figure}[!ht]
\begin{center}
\includegraphics[width=16cm,height=10cm]{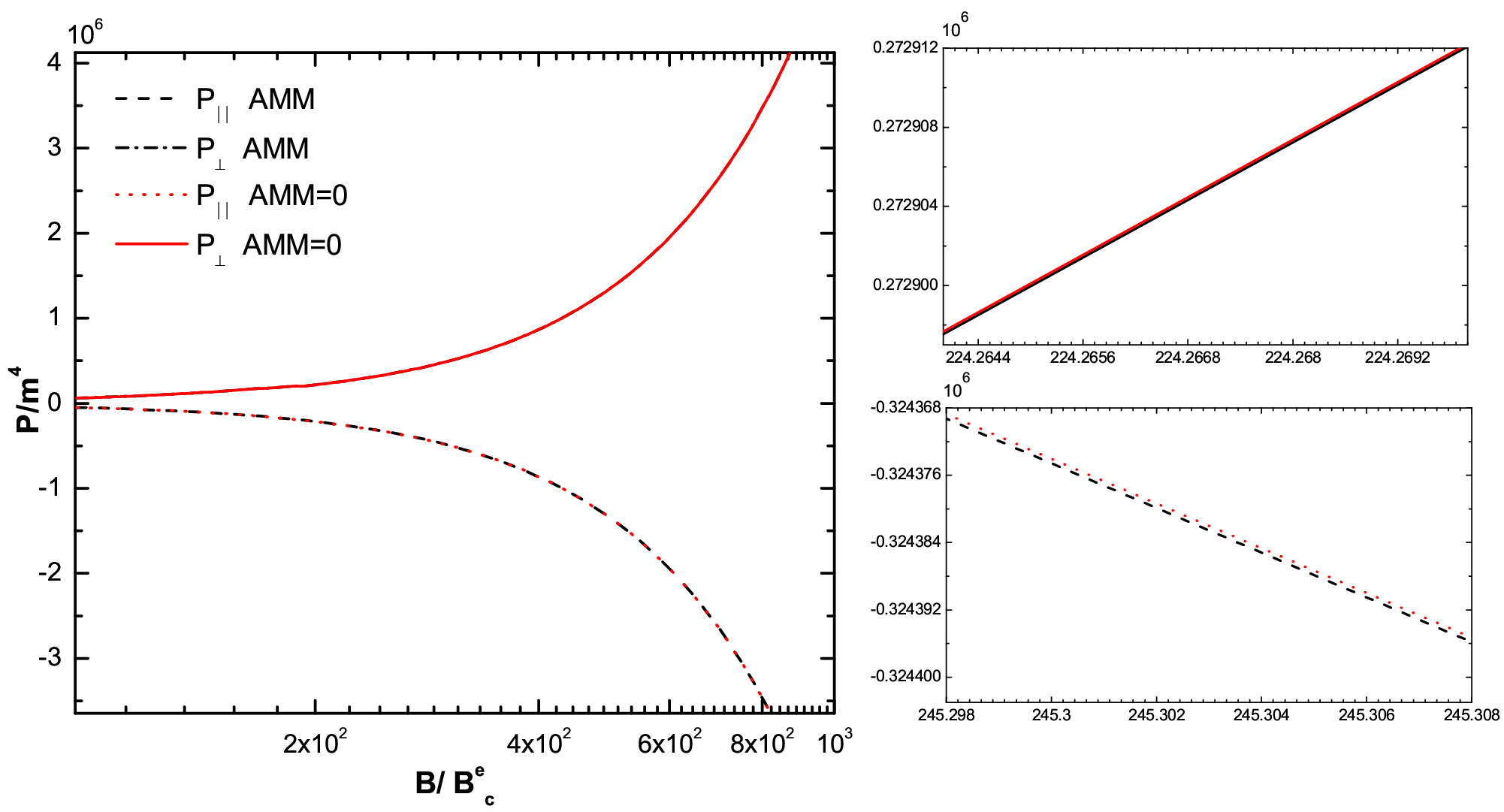}
\caption{\footnotesize (Color online) \emph{Maxwell term effects on the pressures.} Profile of the parallel and transverse pressures versus the magnetic field in the strong-field region including vacuum and Maxwell contributions and using $\mu=10.0$~MeV and $m/\mu\simeq0.051$.  Black curves indicate $AMM\neq0$, red curves indicate $AMM=0$. The Maxwell term totally changes the profile of the pressures, as can be seen by comparing these plots with those shown in Figs. (\ref{fig3}) and (\ref{fig4}). At sufficiently high $B$, the Maxwell contribution, which enters with different signs on each pressure, has a magnitude large enough to produce a noticeable decrease (increase) in the parallel (transverse) pressure. The zooms in the right-hand panels show how the Maxwell pressure practically erases the already small differences between the pressure with and without the AMM.} \label{fig7}
\end{center}
\end{figure}

The transverse pressure in Figs.\ref{fig3} and \ref{fig4} displays a de Haas-van Alphen oscillation associated with the transition of the fermion filling from the Landau level $l=1$ to the LLL as the magnetic field increases. De Haas-van Alphen oscillations are also present in the energy density (Fig.(\ref{fig5})) and in the magnetization (Fig.\ref{fig6}). The size of the de Haas-van Alphen oscillations are not large enough in the parallel pressure to be observable at the scales under consideration. It is also evident from Figs.\ref{fig5} and \ref{fig6} that the AMM makes no significant contribution to the energy density or the magnetization.

\begin{figure}[!ht]
\begin{center}
\includegraphics[width=12cm,height=10cm]{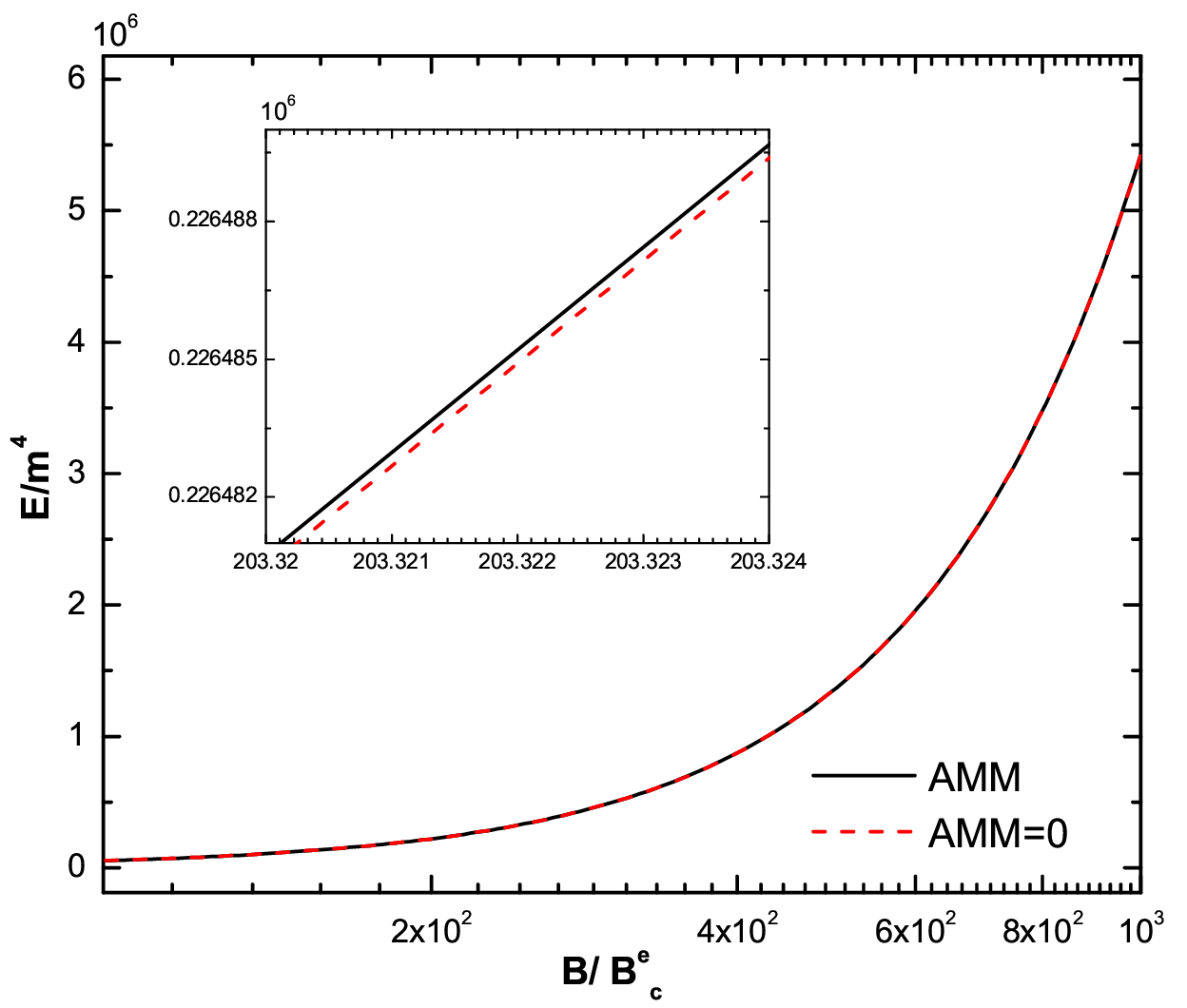}
\caption{\footnotesize (Color online)  \emph{Maxwell term effects on the energy density.} Energy density versus the magnetic field in the strong-field region including vacuum and Maxwell contributions and using $\mu=10.0$~MeV and $m/\mu\simeq0.051$. Black (red) curve indicates $AMM\neq0$ ($AMM=0$). The Maxwell term completely changes the behavior of the energy density with the field, as can be gathered by comparing these plots with those of Fig.(\ref{fig5}). The zoom shows that there are no substantial differences when the AMM is present.} \label{fig8}
\end{center}
\end{figure}

The effects of the Maxwell term on the pressures in the strong field region are explored in Fig.\ref{fig7}. Without the $B^2/2$ term, the parallel pressure increases with the field (see Figs.\ref{fig3} and \ref{fig4}). However, once this contribution is added, it wins over the rest of the terms in the thermodynamic potential and the field-dependence is inverted, eventually becoming negative for strong enough fields. For the transverse pressure, the opposite effect occurs. On the other hand, the Maxwell contribution completely erases any difference produced by the AMM among the parallel and transverse pressures. In a similar fashion, a Maxwell term modifies the profile of the energy density with the field in the strong-field region, making it to increase with the field as seen in Fig.\ref{fig8}. All these results underline that once the field becomes of the order or larger than the chemical potential, the Maxwell term dominates the behavior of all the thermodynamical quantities.

\begin{figure}[!ht]
\begin{center}
\includegraphics[width=12cm,height=10cm]{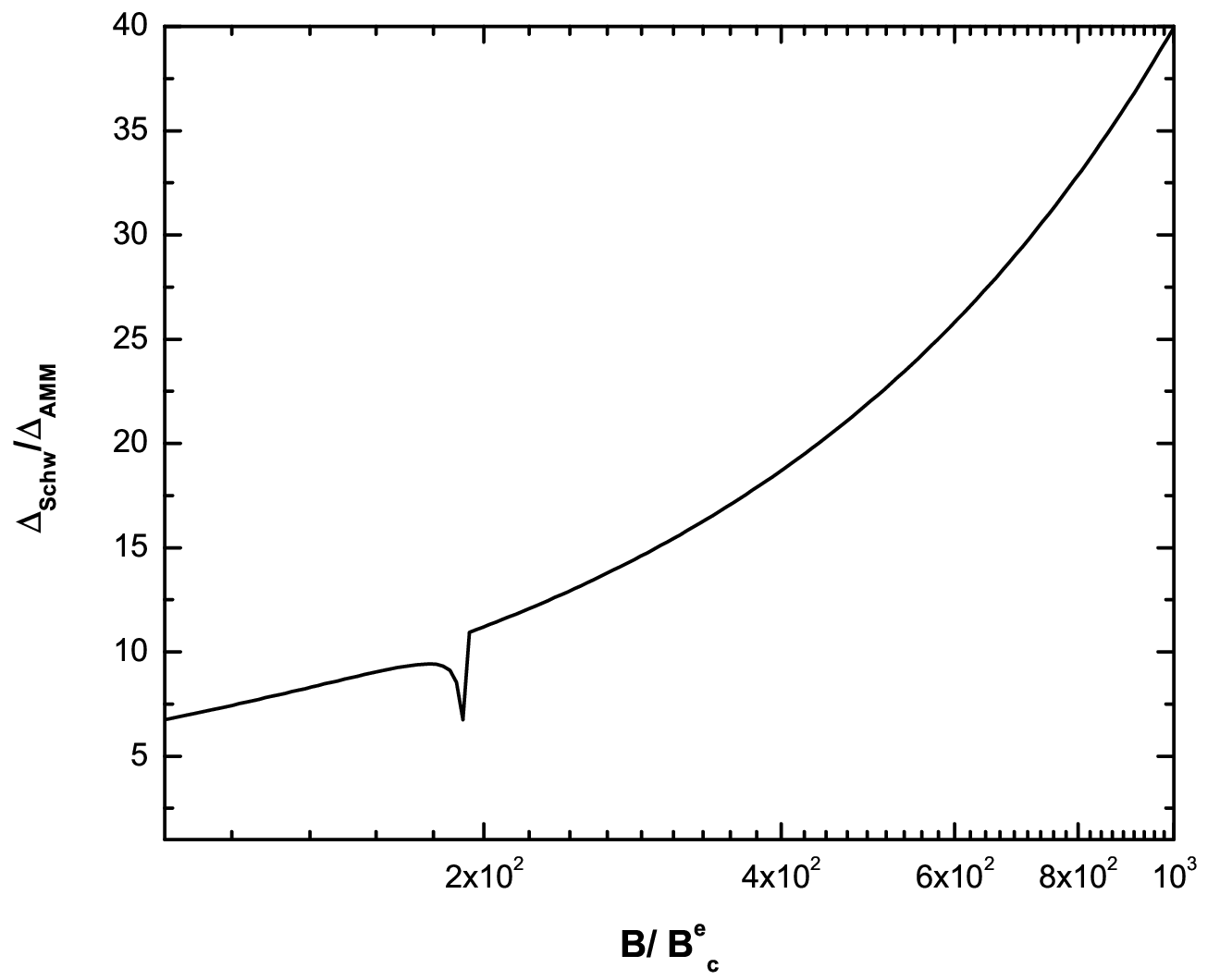}
\caption{\footnotesize \emph{Ratio of the variation of the parallel pressures with AMM's a la Schwinger and in the strong-field limit, $\Delta_{Schw}/\Delta_{AMM}=|P_\|(Schw)-P_\|(AMM=0)|/|P_\|(AMM)-P_\|(AMM=0)|$, for magnetic fields in the strong-field regime}. Notice that the consequence of erroneously using the Schwinger AMM in the strong-field approximation is to change the contribution of the AMM to the parallel pressure in one order of magnitude. The kink corresponds to the  field value where the transition of the fermion filling from the first LL to the LLL takes place.} \label{fig9}
\end{center}
\end{figure}

Now, to show the implications of erroneously using the Schwinger formula for the AMM (\ref{anomalosandmass2}) in the strong-field approximation, let us substitute $\mathcal{T}_1$ in (\ref{T-0}) with $\mathcal{T}=({\alpha}/{2\pi})({eB}/{2m})$ and find the parallel pressure including the vacuum term, which we denote as $P_\|(Schw)$. Let us define now the ratio

\begin{equation}
\frac{\Delta_{Schw}}{\Delta_{AMM}}=\frac{|P_\|(Schw)-P_\|(AMM=0)|}{|P_\|(AMM)-P_\|(AMM=0)|}
\end{equation}
where $P_\|(AMM)$ is the parallel pressure defined from (\ref{T-0}) with $\mathcal{T}_1$ taken from (\ref{T1}), and $P_\|(AMM=0)$ with $\mathcal{T}_1=0$.

\begin{figure}[!ht]
\begin{center}
\includegraphics[width=12cm,height=10cm]{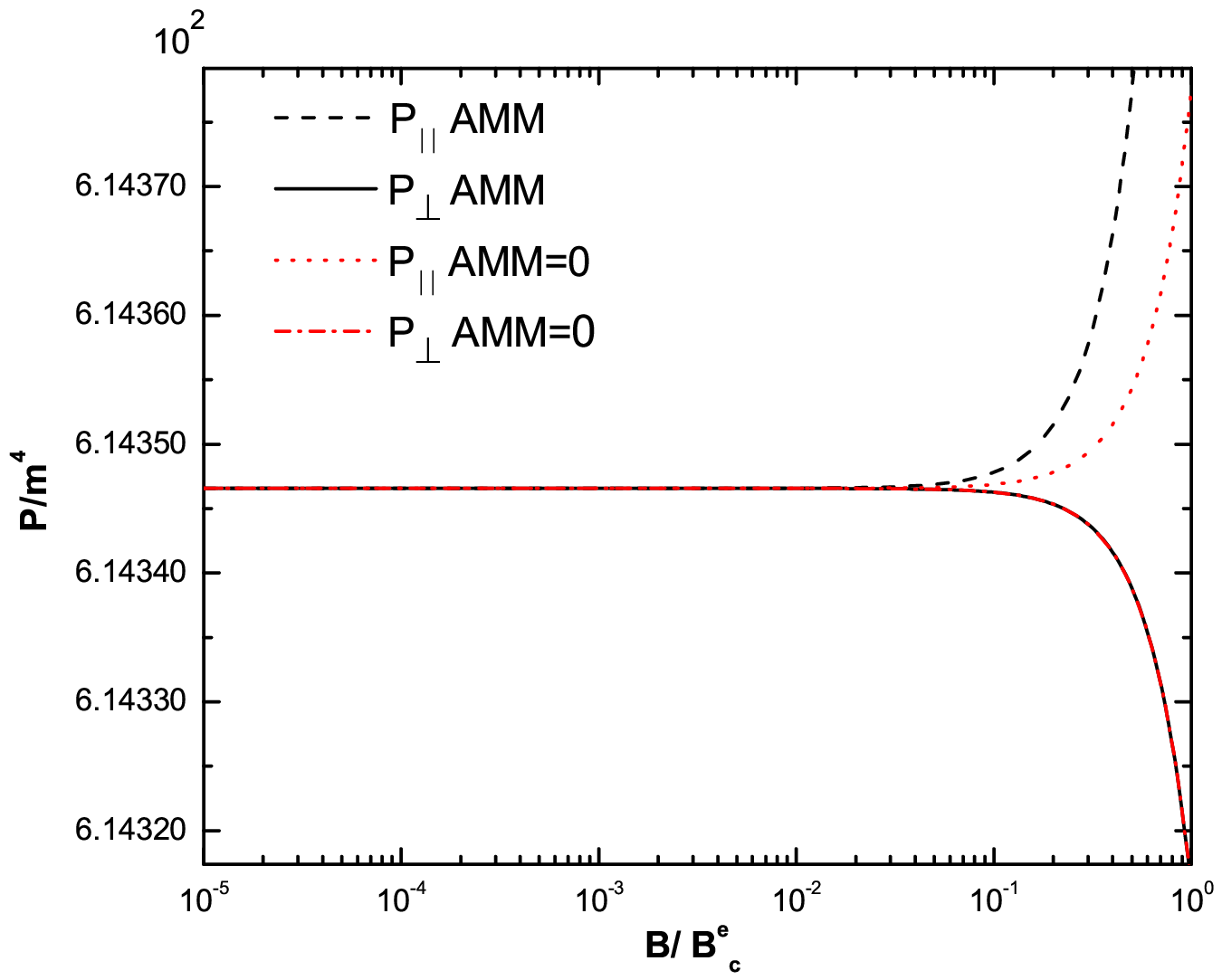}
\caption{\footnotesize  (Color online) \emph{AMM effects on the parallel and transverse pressures in the weak field region without the Maxwell term}. Profile of the parallel and transverse pressures with the magnetic field in the weak-field region including vacuum contribution, but ignoring the Maxwell pressure. The values $\mu=10.0$~MeV and $m/\mu\simeq0.051$ were used. Black (red) curve indicates $AMM\neq0$ ($AMM=0$). Notice that the AMM does not produce any significant effect on the pressures. In the region of interest the two pressures remain positive.} \label{fig10}
\end{center}
\end{figure}

From Fig.\ref{fig9} we see that the net effect of erroneously using the Schwinger AMM in the strong-field approximation is to change in one order of magnitude the contribution of the AMM to the parallel pressure. That is, although the Schwinger's AMM makes an $\alpha$ contribution into the thermodynamic potential, its linear dependence on the field produces a larger effect at strong field than the square logarithmic behavior found for the AMM in the strong-field approximation in (\ref{T1}). The kink appearing in Fig.\ref{fig9} corresponds to the  field value where the de Haas-van Alphen oscillation takes place in Figs.\ref{fig3} and \ref{fig4}. Thus, for the scale used in Fig. \ref{fig9} the de Haas-van Alphen oscillation of the parallel pressure becomes apparent. The effect of the Schwinger's AMM is to stiffer both pressures. Nevertheless, when the Maxwell contributions are added one can see that at such high fields the effect of the Schwinger's AMM also becomes negligible. In this analysis we were considering only the parallel pressure to make contact with other results in the literature, because the parallel pressure is what has been usually considered when the pressure splitting has been ignored, but if we repeat the calculation for the transverse pressure the result will be identical.

\subsection{EoS in the weak-limit approximation}

\begin{figure}[!ht]
\begin{center}
\includegraphics[width=12cm,height=10cm]{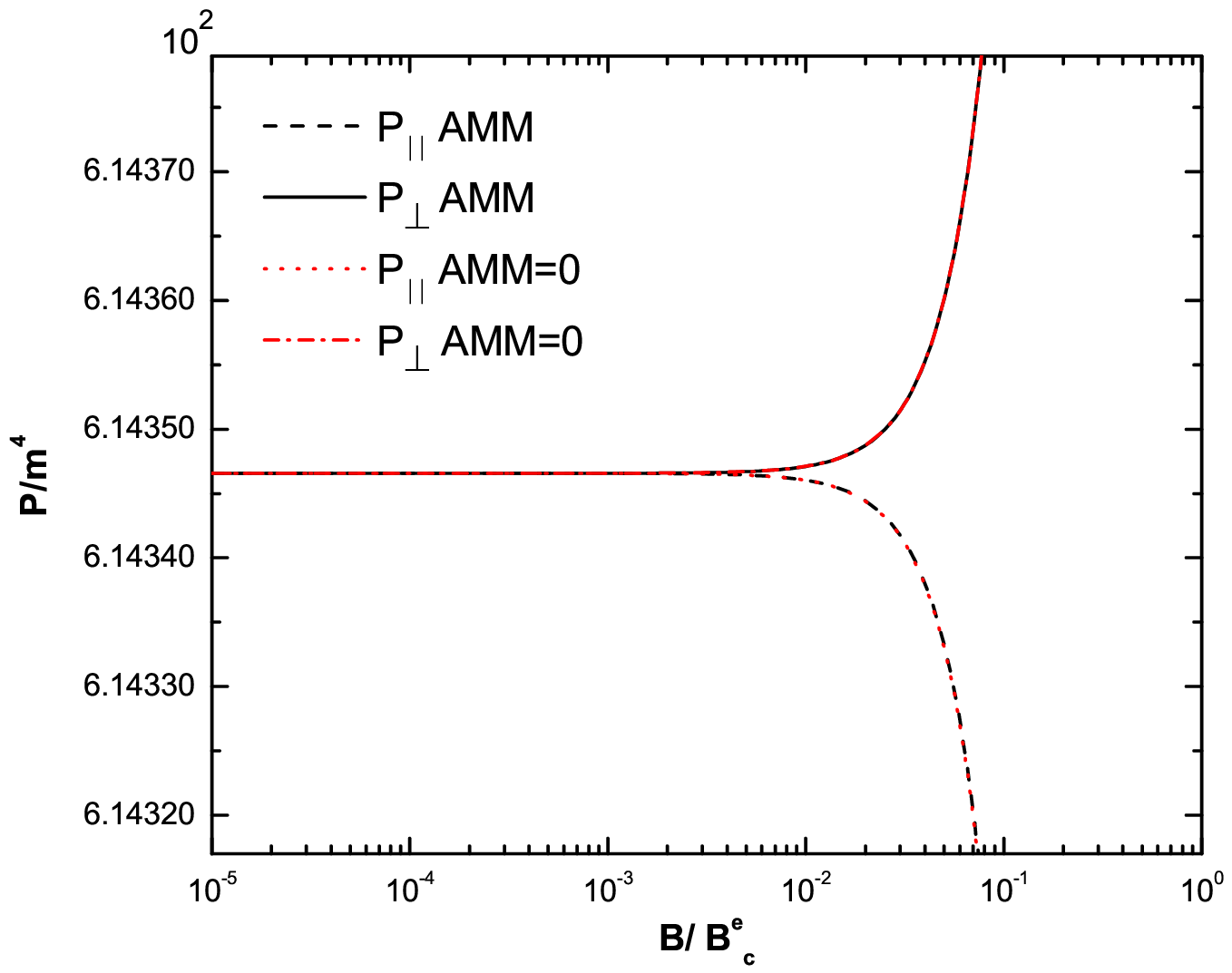}
\caption{\footnotesize  (Color online) \emph{Effects of the Maxwell term on the parallel and transverse pressures in the weak-field region.} Profile of the parallel and transverse pressures with the field in the weak-field region including vacuum and Maxwell contributions. The values $\mu=10.0$~MeV and $m/\mu\simeq0.051$ were used. Black (red) curve indicates $AMM\neq0$ ($AMM=0$). Due to the Maxwell contribution the transverse pressure increases with the field, while the parallel pressure decreases, but it does not cross zero. As in the strong-field case, the AMM effects are totally erased by the Maxwell contribution.} \label{fig11}
\end{center}
\end{figure}

In the weak-field approximation, the thermodynamic potential at zero temperature is given by the sum of the renormalized leading vacuum term (\ref{poteffRen-2}) and the medium contribution (\ref{TP-munew})
\bea \label{TP-weak}
   \Omega^{W(R)}(T=0)&=&\Omega^{W(R)}_{vac}+\Omega^W_\mu=\frac{1}{8\pi^2}
    \left\{\frac{(eB)^4}{45m^4}
         +\frac{(eB)^2\mathcal{T}^2}{3m^2} \right\}
\nonumber \\
&&-\frac{1}{8\pi^2}
 \Bigg\{\frac{2}{3}\mu p_F^3-m^2\mu p_F
       +m^4\ln\left(\frac{p_F+\mu}{m}\right)\Bigg\}\theta(\mu-m)
\nonumber \\
    &&-\frac{\mathcal{T}^2}{4\pi^2}\left(\mu p_F+m^2\ln\left(\frac{p_F+\mu}{m}\right) \right)\theta(\mu-m)
\nonumber \\
   &&
   +\frac{|eB|m\mathcal{T}}{2\pi^2}\ln\left(\frac{p_F+\mu}{m}\right)\theta(\mu-m)
    -\frac{|eB|^2}{12\pi^2}\ln\left(\frac{p_F+\mu}{m}\right)\theta(\mu-m)
\eea
Here the AMM $\mathcal{T}$ is taken in the Schwinger approximation (\ref{anomalosandmass2}).

We can now use (\ref{TP-weak}) to calculate the parallel and transverse pressures in the weak-field region. We plot them, ignoring the Maxwell term, as functions of the magnetic field in Fig.\ref{fig10}. For comparison, we also plotted the corresponding pressures with no AMM. It can be gathered from the figure that the AMM fails to produce a significant effect on the pressures also in the weak field case ($B\lesssim B_c^e/10$). This is easy to understand since at weak field, $\mathcal{T}\sim \alpha (eB/2m)$ makes a small correction to the thermodynamic potential. In contrast to the strong-field regime results, in the weak-field range ($eB < m^2<\mu^2$) the two pressures remain positive.

Similarly to what happens at strong fields, once the Maxwell pressure is included any effect of the AMM is erased and the dependence of the pressures on the magnetic field is inverted with respect to the case without it. As shown in Fig.\ref{fig11}, with the Maxwell contribution inserted, the transverse pressure increases with the field, while the parallel pressure decreases. However, the latter decreases slowly and does not become negative.

We can express the percentage splitting between the parallel and transverse pressures in the weak-field region as
\begin{equation}\label{Pressure-Splitting-ratio}
\Delta[\%]=\frac{\mid{P_\bot}-{P_\parallel}\mid}{\mid{P_\parallel}(eB\sim0)\mid}\times 100
\end{equation}
All the quantities in this formula contain the Schwinger AMM denoted by $\mathcal{T}$. As can be gathered from Figs.\ref{fig9} and \ref{fig10}, at very weak fields ($B \ll B_c^e$) the two pressures coincide, so we have denoted them in this region by a common symbol $P_\parallel(eB\sim0)$).

\begin{figure}[!ht]
\begin{center}
\includegraphics[width=12cm,height=10cm]{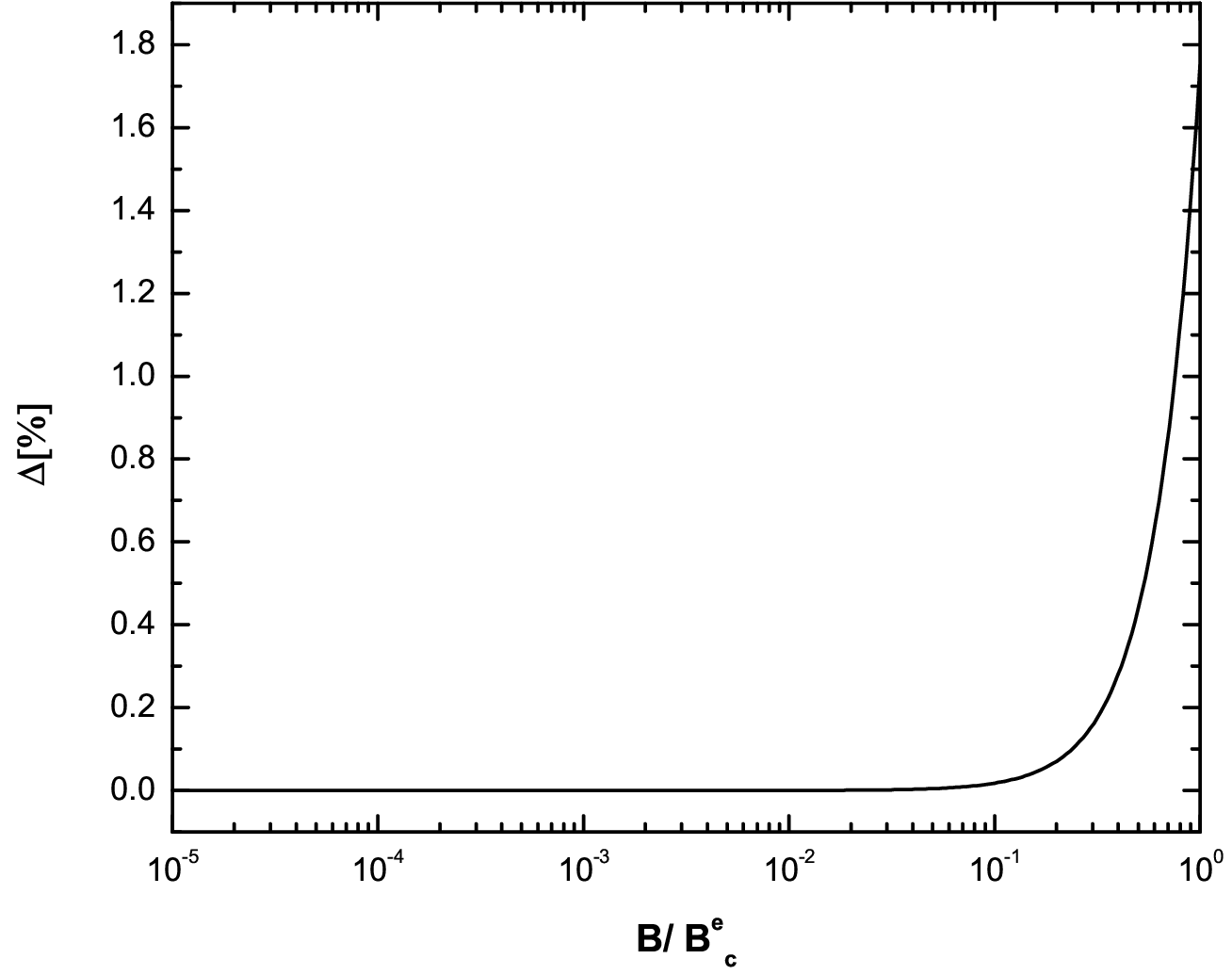}
\caption{\footnotesize \emph{Profile of the pressure splitting percentage versus the magnetic field.} Pressure splitting (Eq. (\ref{Pressure-Splitting-ratio})) vs magnetic field including the Maxwell pressure for $\mu=10.0 MeV.$ Notice that the splitting percentage is smaller than 2, so in the weak-field regime the pressure splitting is negligibly small.} \label{fig13}
\end{center}
\end{figure}

A plot of the percentage splitting (\ref{Pressure-Splitting-ratio}) as a function of the magnetic field is given in Fig.\ref{fig13} using $\mu=10.0$~MeV. Here we included the Maxwell contribution. We can see that for fields up to $B\sim B_c^e$ the splitting is much smaller than $2\%$. For strong fields it can be ten times larger \cite{EOS-Ferrer}. Thus, we conclude that in the weak-field regime the pressure splitting is negligibly small and the inclusion of the AMM makes no difference.

\section{Conclusions}\label{section8}

In this paper we presented a thorough study of the EoS of charged fermions endowed with AMM in a dense medium. The calculations were done in the strong and weak field limits, including on each case, the corresponding AMM previously obtained in those approximations.

To obtain the AMM in the strong and weak field limits we took advantage of the Ritus's approach, which allows diagonalizing the self-energy in momentum space and separating the different LL contributions. We found that at strong field, the particle gets an AMM that depends on the LL's, with a noticeable difference between the value of the AMM for $l=1$, which is proportional to the square logarithm of the magnetic field and equal to half the rest energy of the LLL,  and the AMM for the remaining levels $l>1$, which is much smaller and decreases very rapidly with the field. In sharp contrast, in the weak-field approximation the AMM is independent of the LL and grows linearly with the field.

We calculated the different contributions to the thermodynamic potential: vacuum, medium and Maxwell terms, and found how they affected the energy density and the parallel and transverse pressures in the presence of a magnetic field. We then investigated the role on the EoS of each of these contributions, with and without the inclusion of the particle's AMM in the strong, as well as in the weak-field approximations.

Our main conclusion is that the AMM of charged fermions makes no difference in the EoS, neither at strong, nor at weak magnetic fields. This statement contradicts some claims in the literature \cite{EoS-AMM} about the significant effects of the AMM at strong fields. We believe that the origin of the discrepancy is connected to using Schwinger's result to describe the AMM of all the fermion species entering in the theory and assuming this is a good approach for arbitrary values of the magnetic field. Such an inconsistent treatment to study the effects of the AMM on the thermodynamics of the system is a common element of several works of Ref. \cite{EoS-AMM}, so the claims about the important role of the AMM in the EoS should be taken with a grain of salt, at least for the cases where the claims are connected to the role of the AMM of charged fermions, since our results, and hence our criticism, applies only to the effects of the AMM of charged fermions. It would be interesting in the near future to undertake a similar study about the effect of the AMM of neutral composite fermions.

\acknowledgments%\bigskip
EJF and VI work was supported in part by DOE Nuclear theory grant DE-SC0002179.
The work of APM and DMP has been supported under the
grant of the ICTP Office of External Activities through NET-35.
APM acknowledges the hospitality and support of the International
Center for Relativistic Astrophysics Network (ICRAnet) and Instituto
de Ciencias Nucleares-UNAM, Mexico DF, where part of this work was developed. DMP thanks CLAF-ICTP for the support received during the realization of this work. APM and DMP thanks A. Cabo and H. Perez Rojas for discussions.

\appendix
\section{Thermodynamic potential in the strong-field approximation}\label{appendixA-1}

\subsection{Vacuum contribution}

We are interested here in the vacuum contribution to the thermodynamic potential in the strong field region  $\Omega^S_{vac}$. It is obtained from $\Omega_{vac}$ in (\ref{omega-2}), after taking into account that at strong field all $\mathcal{T}_l$ with $l>1$ can be neglected and hence the energy modes $ E_{\eta\sigma l}$ become
\begin{equation}\label{spectra}
E_{\pm+0}=\pm\sqrt{p_3^2+m^2}, \quad E_{\pm\sigma1}=\pm\sqrt{p_3^2+(\sqrt{m^2+2eB}+\sigma \mathcal{T}_1)^2}, \quad E_{\pm\sigma (l>1)}=\pm\sqrt{p_3^2+m^2+2eBl}
\end{equation}

Therefore,
\bea \label{omega-vac}
 \Omega^S_{vac}&=&- \frac{eB}{8\pi^2} \int_{-\infty}^{\infty}  dp_3  \sum_{\eta\sigma l} \vert E_{\eta\sigma l}\vert \nonumber
 \\
 &=&- \frac{eB}{4\pi^2} \int_{-\infty}^{\infty}  dp_3[\sqrt{p_3^2+m^2}+\sum_\sigma \sqrt{p_3^2+(\sqrt{m^2+2eB}+\sigma \mathcal{T}_1)^2}
 \nonumber\\
 \qquad &+&2\sum_{l=2}^\infty \sqrt{p_3^2+m^2+2eBl}]
\eea

If we add and subtract $2\sqrt{p_3^2+m^2+2eB}$, it can be written as
\bea\label{vac-2}
\Omega^S_{vac}&=&- \frac{eB}{4\pi^2} \int_{-\infty}^{\infty}  dp_3 \bigg [ \sum_{\sigma} \sqrt{p_3^2+(\sqrt{m^2+2eB}+\sigma \mathcal{T}_1)^2}-2 \sqrt{p_3^2+m^2+2eB}
\nonumber\\
\qquad &+&\sum_{l=0}^\infty (2-\delta_{l0})\sqrt{p_3^2+m^2+2eBl} \bigg ]
\eea

Performing the sum in LL's, integrating in momentum, and introducing the proper-time representation, we obtain
\bea\label{vac-3}
\Omega^S_{vac}&=& \frac{eB}{8\pi^2} \int_{1/\Lambda^2}^{\infty}  \frac{ds}{s^2} \bigg [ e^ {-s(\sqrt{m^2+2eB}+ \mathcal{T}_1)^2}+e^ {-s(\sqrt{m^2+2eB}-\mathcal{T}_1)^2}
\nonumber\\
&-&2e^ {-s(m^2+2eB)}+e^ {-sm^2}\coth(eBs) \bigg ]
\eea

The terms without the $coth(eBs)$ can be integrated using the formula \cite{prudnikov}
\begin{equation}\label{form}
\int \frac{e^{ax}}{x^2}dx=-\frac{e^{ax}}{x}+aEi(ax),
\end{equation}
where $Ei(ax)$ is the exponential integral function. Evaluating in the limits of integration and using the formula
\begin{equation}\label{form-2}
Ei(ax)=\ln|ax|+\sum_{k=1}^\infty \frac{(ax)^k}{k! k}
\end{equation}
which is valid for $|ax|<1$, we obtain that the leading contribution in the limit of large $\Lambda$ is
\bea\label{vac-int}
 \int_{1/\Lambda^2}^{\infty}  \frac{ds}{s^2} \bigg [ e^ {-s(\sqrt{m^2+2eB}+ \mathcal{T}_1)^2}+e^ {-s(\sqrt{m^2+2eB}-\mathcal{T}_1)^2}-2e^ {-s(m^2+2eB)}\bigg ] \qquad
 \nonumber\\
\simeq 4\mathcal{T}_1^2+2\mathcal{T}_1^2 \ln \left[ \frac{m^2}{\Lambda^2}\right]+ 2\mathcal{T}_1^2 \ln \left[ \frac{2eB}{m^2}\right] \qquad \quad \quad \quad
 \eea
The term proportional to $\coth (eBs)$ in (\ref{vac-3}) has ultraviolet divergencies that can be identified by expanding it in powers of s. They reduce to two terms,
 \begin{equation}\label{form-3}
 \frac{1}{8\pi^2}\int_{1/\Lambda^2}^{\infty}  \frac{ds}{s^3} e^ {-sm^2},
 \end{equation}
and
\begin{equation}\label{form-4}
\frac{1}{8\pi^2} \int_{1/\Lambda^2}^{\infty}  \frac{ds}{s^3} e^ {-sm^2}\frac{(eBs)^2}{3},
 \end{equation}
To eliminate these divergences, one can add and subtract (\ref{form-3}) and (\ref{form-4}) to (\ref{vac-3}) \cite{Proper-Time}. The divergence (\ref{form-3}), which depends on the renormalized mass, is then incorporated into a renormalized vacuum constant, while (\ref{form-4}) is absorbed into the bare Maxwell energy to define renormalized field and charge. Then mass, charge and magnetic field are all renormalized in the standard way. The second term in the rhs of (\ref{vac-int}) is divergent though. Keeping in mind that the parameter $\mathcal{T}_1$ is finite, since it is given by the AMM found from the renormalized one-loop self-energy of the original theory, hence was introduced in the effective theory as a given finite parameter, we can treat this term in the same way done in \cite{Proper-Time} with (\ref{form-3}) and incorporate it into the renormalized vacuum constant.

Therefore, the renormalized vacuum contribution to the thermodynamic potential in the strong field limit takes the form
\bea\label{vac-Ren}
\Omega_{vac}^{S(R)}&=& \frac{1}{8\pi^2}\int_{1/\Lambda^2}^{\infty} ds \frac{e^{-sm^2}}{s^3}[eBs\coth(eBs)-1- \frac{(eBs)^2}{3}]+ \frac{eB}{4\pi^2}\mathcal{T}_1^2\left[\ln \frac{2eB}{m^2}+2\right]
\eea

To extract the strong-field contribution to the integral in (\ref{vac-Ren}), one separates it in two pieces
\begin{equation}\label{integ}
 \int_{1/\Lambda^2}^{\infty}ds= \int_{1/\Lambda^2}^{1/eB}ds+\int_{1/eB}^{\infty}ds
\end{equation}
The first integral can be done expanding the $coth$ in powers of its argument and gives a negligible contribution in the limit $\Lambda \to \infty$. The second integral is finite. Its leading contribution is given by
$-\frac{\alpha B^2}{6\pi}\ln\left(\frac{eB}{m^2}\right)$, which coincides with the strong-field result found in \cite{Raga}.
 Then, keeping just the leading-in-$\alpha$ terms in the strong-field approximation, the renormalized vacuum potential reduces to
\bea\label{vac-Ren-22}
\Omega_{vac}^{S(R)}=-\frac{\alpha B^2}{6\pi}\ln\left(\frac{eB}{m^2}\right)+\frac{eB}{4\pi^2} \mathcal{T}_1^2\left[\ln \frac{2eB}{m^2}+2\right]
\eea
Substituting $\mathcal{T}_1$ with the result (\ref{omega-3}), we obtain
\bea\label{vac-Ren-2}
\Omega_{vac}^{S(R)}=-\frac{\alpha B^2}{6\pi}\ln\left(\frac{eB}{m^2}\right)+\frac{eB}{4\pi^2} \left[\left(\frac{\alpha m}{8\pi}\right)\ln^2{\left(\frac{2eB}{m^2}\right)}\right]^2\left[\ln \frac{2eB}{m^2}+2\right]
\eea
so clearly the leading vacuum contribution is mainly determined by the strong-field contribution at zero AMM \cite{Ragazzon}, given by the first term in the rhs of (\ref{vac-Ren-2}).

\subsection{Finite-density contribution}

At zero temperature the contribution of the fermions in the Fermi sphere to the thermodynamic potential is given by
\bea \label{omega-mu}
    \Omega^S_{\mu}= - \frac{eB}{8\pi^2} \int_{-\infty}^{\infty}  dp_3  \sum_{\eta \sigma l}( \vert E_{\eta \sigma l}-\mu \vert-\vert E_{\eta \sigma l} \vert )
\eea
with energy spectrum given in (\ref{spectra}).

Summing in $\eta$ we get
\bea \label{omega-mu-2}
    \Omega^S_{\mu}=& -& \frac{eB}{8\pi^2} \sum_{\sigma l}\int_{-\infty}^{\infty}  dp_3  \left [\vert E_{+\sigma l}-\mu\vert-2 E_{+\sigma l} + ( E_{+\sigma l}+\mu)  \right ]
    \nonumber\\
    =& - &\frac{eB}{8\pi^2} \sum_{\sigma l}\int_{-\infty}^{\infty}  dp_3   \left [E_{+\sigma l}-\mu+E_{+\sigma l}+\mu-2E_{+\sigma l} \right] \theta(E_{+\sigma l}-\mu)
     \nonumber\\
    & - &\frac{eB}{8\pi^2} \sum_{\sigma l}\int_{-\infty}^{\infty}  dp_3   \left [\mu- E_{+\sigma l}+ E_{+\sigma l}+\mu-2 E_{+\sigma l} \right ] \theta(\mu- E_{+\sigma l})
       \nonumber\\
       =&- &\frac{eB}{4\pi^2} \sum_{\sigma l}\int_{-\infty}^{\infty}  dp_3  \left (\mu- E_{+\sigma l}\right) \theta(\mu- E_{+\sigma l})
\eea

Then, after integration in momentum and summing in spin, we obtain
\bea \label{omega-mu-3}
    \Omega^S_{\mu}=& -& \frac{eB}{4\pi^2}   \theta(\mu-m) \left [\mu\sqrt{\mu^2-m^2} -m^2 \ln \left(\frac{\mu +\sqrt{\mu^2-m^2}}{m}\right)  \right ]
    \nonumber\\
    & - & \frac{eB}{4\pi^2} \sum_{l=1}^\infty \theta(\mu-M_l^+)\left[\mu\sqrt{\mu^2-(M_l^+)^2}-(M_l^+)^2 \ln \left (\frac{\mu+\sqrt{\mu^2-(M_l^+)^2}}{M_l^+}\right)\right ]
     \nonumber\\
    & - & \frac{eB}{4\pi^2} \sum_{l=1}^\infty \theta(\mu-M_l^-)\left[\mu\sqrt{\mu^2-(M_l^-)^2}-(M_l^-)^2 \ln \left (\frac{\mu+\sqrt{\mu^2-(M_l^-)^2}}{M_l^-}\right)\right ]
\eea
with
\begin{equation}\label{magnetic-mass}
M_l^\pm =\sqrt{2eBl+m^2} \pm \mathcal{T}_l, \quad l\geqslant 1
\end{equation}
where we have neglected the radiative corrections to the mass for all LL's, including the LLL. As already mentioned, in the strong-field limit, $\mathcal{T}_l\ll \mathcal{T}_1$ for $l>1$ (see Fig. 1), which allows us to neglect in (\ref{magnetic-mass}) the contribution of $\mathcal{T}_l$ for $l>1$. Hence, (\ref{omega-mu-3}) reduces to
\bea \label{omega-mu-4}
    \Omega^S_{\mu}=& -& \frac{eB}{4\pi^2}   \theta(\mu-m) \left [\mu\sqrt{\mu^2-m^2} -m^2 \ln \left(\frac{\mu +\sqrt{\mu^2-m^2}}{m}\right)  \right ]
    \nonumber\\
    & - & \frac{eB}{4\pi^2}  \theta(\mu-M_1^+)\left[\mu\sqrt{\mu^2-(M_1^+)^2}-(M_1^+)^2 \ln \left (\frac{\mu+\sqrt{\mu^2-(M_1^+)^2}}{M_1^+}\right)\right ]
     \nonumber\\
    & - & \frac{eB}{4\pi^2}  \theta(\mu-M_1^-)\left[\mu\sqrt{\mu^2-(M_1^-)^2}-(M_1^-)^2 \ln \left (\frac{\mu+\sqrt{\mu^2-(M_1^-)^2}}{M_1^-}\right)\right ]
    \nonumber\\
     & - & \frac{eB}{2\pi^2} \sum_{l=2}^\infty \theta(\mu-M_l)\left[\mu\sqrt{\mu^2-(M_l)^2}-(M_l)^2 \ln \left (\frac{\mu+\sqrt{\mu^2-(M_l)^2}}{M_l}\right)\right ]
\eea
where
\begin{equation}\label{magnetic-mass-2}
M_l=\sqrt{2eBl+m^2} , \quad l>1
\end{equation}
Expression (\ref{omega-mu-4}) coincides with Eq. (\ref{omega-0}).

\subsection{Finite-temperature contribution}

The contribution at finite temperature is given by
\bea \label{omega-T}
    \Omega^S_{\beta}&=&-\frac{eB}{4\pi^2} \int_{-\infty}^{\infty}  dp_3  \sum_{\eta\sigma l} \frac{1}{\beta} \ln(1+e^{-\beta\vert E_{\eta\sigma l}  -\mu\vert})
\eea

After summing in $\eta$ and $\sigma$, and taking into account the corresponding spectra (\ref{spectra}), we obtain
\bea
\Omega^S_\beta=&-&\frac{eB}{2\pi^2\beta} \int_{0}^{\infty}dp_3\ln(1+e^{-\beta|\sqrt{p_3^2 +m^2}+\mu|})(1+e^{-\beta|\sqrt{p_3^2 +m^2}-\mu|})
    \nonumber
   \\
&-&\frac{eB}{2\pi^2\beta}\int_{0}^{\infty}dp_3\ln(1+e^{-\beta|\sqrt{p_3^2 +{M_1^+}^2}+\mu|})(1+e^{-\beta|\sqrt{p_3^2 +{M_1^+}^2}-\mu|}) \nonumber
 \\
 &-&\frac{eB}{2\pi^2\beta} \int_{0}^{\infty}dp_3\ln(1+e^{-\beta|\sqrt{p_3^2 +{M_1^-}^2}+\mu|})(1+e^{-\beta|\sqrt{p_3^2 +{M_1^-}^2}-\mu|})\nonumber
  \\
&-&\frac{eB}{\pi^2\beta}\sum_{l=2}^{\infty}\int_{0}^{\infty}dp_3\ln(1+e^{-\beta|\sqrt{p_3^2 +{M_l}^2}+\mu|})(1+e^{-\beta|\sqrt{p_3^2 +{M_l}^2}-\mu|}),
 \label{omega-beta-2}
\eea
Here we are using the notation given in (\ref{magnetic-mass}) and (\ref{magnetic-mass-2}).

\section{Thermodynamic potential \'a  la Dittrich}\label{appendixD}

In this Appendix, we will discuss the steps followed in \cite{DittrichAMM} to calculate the vacuum contribution of the thermodynamic potential for charged fermions endowed with AMM in a magnetic field. In doing that, we will call attention to some issues we have found in those calculations that explain the difference between the strong-field results (\ref{vac-Ren-App}) and (\ref{Dittrich}).

Let us start by writing Eq.(2.1) of Ref.~\cite{DittrichAMM}
\bea
   (m-\frac{\mu}{2}\sigma_{\mu\nu}F^{\mu\nu}+\gamma\cdot\Pi)G(x,y;A)=\delta(x-y)
\label{dittrich1}
\eea
Taking into account that the fermion propagator can be written in term
of a scalar propagator, $\Delta(x,y;A)$, as
\bea
G(x,y;A)=(m-\frac{\mu}{2}\sigma_{\mu\nu}F^{\mu\nu}-\gamma\cdot\Pi)\Delta(x,y;A),
\label{dittrich2}
\eea
then, we can write \Eq{dittrich1} in term of $\Delta(x,y;A)$, as
\bea
   \left[(m-\frac{\mu}{2}\sigma_{\mu\nu}F^{\mu\nu}+\gamma\cdot\Pi)(m-\frac{\mu}{2}\sigma_{\mu\nu}F^{\mu\nu}-\gamma\cdot\Pi)\right]\Delta(x,y;A)=\delta(x-y),
\eea
which, with the help of the identity $
   \left[\gamma_\rho,\sigma_{\mu\nu}\right]F^{\mu\nu}\Pi^\rho
    =4i\gamma^\mu F_{\mu\rho}\Pi^\rho$
, simplifies to
\bea
   \left[\left(m-\frac{\mu}{2}\sigma_{\mu\nu}F^{\mu\nu}\right)^2-(\gamma\cdot\Pi)^2
  {-2\mu i\gamma^\mu F_{\mu\nu}\Pi^\nu}
  \right]\Delta(x,y;A)=\delta(x-y)
\label{dittrich2a}
\eea
By comparing \Eq{dittrich2a} with Eq.(2.3) of Ref.~\cite{DittrichAMM},
we can see that the last term inside the squared parenthesis in (\ref{dittrich2a}) is missed in Ref.~\cite{DittrichAMM}. On the other hand, if we compare the quasiparticle spectrum that corresponds to Eq.(2.3) of Ref.~\cite{DittrichAMM}
\begin{equation}\label{Dittrich-spectrum}
p_0^2=p_3^2+2leB+[m-\mathcal{T}]^2,
\end{equation}
with the very well-known one for charged particles with magnetic
moment in a uniform magnetic field, we have that the  spectrum (\ref{Dittrich-spectrum}) can be found from  (\ref{varepsilondef1new})  in the limit
\begin{equation}\label{spectrum-limit}
\sqrt{1+\frac{2eBl}{m^2}}\simeq 1,
\end{equation}
which will be only valid at very weak field. Then, the strong-field result (\ref{Dittrich}), which was found after the consideration (\ref{spectrum-limit}), is not valid.

To compare our thermodynamical potential with the one reported in \cite{DittrichAMM}, let us rewrite our  \Eq{poteff7} (see also \Eq{poteff85}) as
\bea
   {\Omega}_{vac}^W &=&- \frac{|eB|}{2(2\pi)^2}
   \int_0^\infty
   \frac{ds}{s^2}\sum_{l=0}^{\infty}(2-\delta_{l0})
   e^{-s\left(m^2+p_\perp^2+\mathcal{T}^2\right)}
   \cosh\left(2s\mathcal{T}\sqrt{m^2+p_\perp^2}\right)
\nonumber \\
&&+\frac{|eB|}{2(2\pi)^2}
    \int_0^\infty\frac{ds}{s^2}
    e^{-s\left(m^2+\mathcal{T}^2\right)}
    \sinh(2s\mathcal{T}m)
\label{dittrichpot1}
\eea

Applying in \Eq{dittrichpot1} the approximation (\ref{spectrum-limit}), we obtain
\bea
   {\Omega}_{vac}^W &\approx&- \frac{|eB|}{2(2\pi)^2}
   \int_0^\infty
   \frac{ds}{s^2}\sum_{l=0}^{\infty}(2-\delta_{l0})
   e^{-s\left(m^2+p_\perp^2+\mathcal{T}^2\right)}
   \cosh\left(2s\mathcal{T}m\right)
\nonumber \\
&&+\frac{|eB|}{2(2\pi)^2}
    \int_0^\infty\frac{ds}{s^2}
    e^{-s\left(m^2+\mathcal{T}^2\right)}
    \sinh(2s\mathcal{T}m)
\label{dittrichpot2}
\eea
Once we perform the summation over all Landau levels, we arrive at
\bea
   \Omega_{vac}^W &=& -\frac{|eB|}{2(2\pi)^2}
   \int_0^\infty
   \frac{ds}{s^2}
   e^{-s\left(m^2+\mathcal{T}^2\right)}\left\{
   \coth\left(eBs\right)\cosh\left(2s\mathcal{T}m\right)
   -\sinh{(2s\mathcal{T}m)}\right\}
\label{dittrichpot4}
\eea

In  \cite{DittrichAMM}, expression (\ref{dittrichpot4}) was  simplified even more considering that $\mathcal{T}\ll eB$ and then taking the small-$\mathcal{T}$ limit for the functions        $\cosh$ and $\sinh$ to obtain, after renormalization, the expression (2.10) of Ref. \cite{DittrichAMM},
\bea
   {\Omega} &=& -\frac{1}{2(2\pi)^2}
   \int_0^\infty
   \frac{ds}{s^3}
   e^{-s\left(m^2+\mathcal{T}^2\right)}\left[
   eBs\coth\left(eBs\right)-1-\frac{1}{3}(eBs)^2\right].
\label{dittrichpot5}
\eea
In our calculation, we went beyond the leading term in the expansion at small-$\mathcal{T}$ of the $\cosh$ and $\sinh$ functions, since there are some extra $\mathcal{T}$-dependent terms that contribute to the divergency. Then, in addition to the usual renormalization subtraction carried out in (\ref{dittrichpot5}), we had to subtract  several $\mathcal{T}$-dependent terms (see Eqs. (\ref{poteff8})-(\ref{poteff10})). Nevertheless, in the leading weak-field approximation taken up to $\mathcal{O}((eB)^4)$ order and in the $\alpha^2$ correction, the effective potential (\ref{dittrichpot5}) reduces to the one we found in (\ref{poteffRen}) for the weak-field approximation.

\section{From Ritus to Schwinger propagator}\label{appendixB}

To perform the summation over all LL's, let us
rewrite \Eq{fulllan3b} by using the Schwinger proper time representation
\bea
    G(x,y)&=&-i(\slsh{\Pi_x}+m)
         \int_0^\infty ds\
         \sumint\frac{d^4p}{(2\pi)^4}e^{is(\overline{p}^2-m^2+i\epsilon)}
        \mathbb{E}_{p}(x)\Pi(l)\overline{\mathbb{E}}_{p}(y)
\label{fulllan4}
\eea
Now, writing the Ritus eigenfunctions explicitly in \Eq{fulllan4}, we obtain
\bea
 G(x,y)
&=&-i(\slsh{\Pi_x}+m)\int_0^\infty ds\
         \int\frac{dp_0p_2p_3}{(2\pi)^4}e^{is(p^2_{||}-m^2+i\epsilon)}
          e^{-i[(x_0-y_0)p_0-(x_2-y_2)p_2-(x_3-y_3)p_3]}
\nonumber \\
   &&\times \sum_{l=0}^{\infty} e^{-is (2eBl)}
    \sum_{\sigma}N_n^2 D_{n}(\rho_x)D_{n}(\rho_y)
    \Pi(l)\Delta(\sigma)
\label{fulllan4b}
\eea
Replacing one parabolic cylinder function by its integral
representation in \Eq{fulllan4b}, and after some manipulations, we get
\bea
 G(x,y)&=&-i(\slsh{\Pi_x}+m)\int_0^\infty ds\
         \int\frac{dp_0p_2p_3}{(2\pi)^4}e^{is(p^2_{||}-m^2+i\epsilon)}
          e^{-i[(x_0-y_0)p_0-(x_2-y_2)p_2-(x_3-y_3)p_3]}
\nonumber \\
   &&\times
    \sqrt{2 eB}
    e^{\frac{1}{4}\rho_x^2}\sum_{r=\pm1}
    \int_0^{\infty} dt\ e^{-\frac{1}{2}t^2}
    e^{ir\rho_xt}
    \sum_{l=0}^\infty\sum_{\sigma=\pm1}
   \frac{ \left(-ir t \right)^n e^{-is (2eBl)}}{n!}
    D_{n}(\rho_y)\Pi(l)\Delta(\sigma)
\nonumber \\
\label{beforeidentity}
\eea
Performing the sum over $\sigma$ and using the identity~\cite{bateman}
\bea
   \sum_{n=0}^{\infty} \frac{t^n}{n!} D_n(z)=e^{-\frac{1}{4}z^2+zt-\frac{1}{2}t^2}
\label{identitycomp}
\eea
on each term of \Eq{beforeidentity}, we obtain
\bea
 G(x,y)&=&-i(\slsh{\Pi_x}+m)\int_0^\infty ds\
         \int\frac{dp_0p_2p_3}{(2\pi)^4}e^{is(p^2_{||}-m^2+i\epsilon)}
          e^{-i[(x_0-y_0)p_0-(x_2-y_2)p_2-(x_3-y_3)p_3]}
\nonumber \\
   &&\times
    \sqrt{2 eB}
    e^{\frac{1}{4}(\rho_x^2-\rho_y^2)}
    \int_{-\infty}^{\infty} dt\ e^{-\frac{1}{2}t^2[1-(e^{-i(2eBs)})^2]}
      e^{it[\rho_x-\rho_ye^{-i(2eBs)}]}
\nonumber \\
   &&\times
   \left[\Delta(+)+e^{-is2eB}\Delta(-)
    \right]
\label{summedprop}
\eea
where in the last line we have performed the sum over $r$.

Once we  perform the
integration over $t$ and in momentum $p$, we arrive at the well
known result for the fermion propagator in an external magnetic field
\bea
 G(x,y)
&=&-(\slsh{\Pi_x}+m)
\Phi(x,y)
    \frac{1}{16\pi^2}
    \int_0^\infty \frac{eBds}{s \sin(eBs)}\ e^{-is(m^2-i\epsilon)}
          e^{-i\left\{\frac{1}{4s}(x-y)^2_{||}-\frac{eB}{4\tan(eBs)}(x-y)_{\perp}^2\right\}}
\nonumber \\
    &&\times
    \left[e^{ieBs}\Delta(+)+e^{-ieBs}\Delta(-)
    \right]
\label{summedprop1}
\eea
where
\bea
\Phi(x,y)=e^{i\frac{eB}{2}(x_2-y_2)(x_1+y_1)}.
\label{phase}
\eea
is the well known Schwinger phase, which encodes all gauge dependence of
the fermion propagator (see Ref.~\cite{Chodos}).

It is easy to prove that for the particular gauge ${\bf A}({\bf z})=B
z_1 \hat{\bf z}_2$ and integrating  along a straight-line path from
$y$ to $x$ the Schwinger phase factor can be written as
\bea
   \frac{eB}{2}(x_2-y_2)(x_1+y_1)=\int^{x}_{y} d {\bf z}\cdot {\bf A}({\bf z})
\eea

\section{Thermodynamic potential in the weak-field approximation}\label{appendixC}

\subsection{Vacuum}

By using the proper-time representation, we rewrite \Eq{Omega-vac}  as
\bea
   \Omega_{vac}^W&=&- \frac{|eB|}{2}\sum_{\sigma=\pm1}
   \int_0^\infty
   \frac{ds}{s}
    \int\frac{d^2p_{||}}{(2\pi)^3}
   \left\{
        e^{-is\left\{p_{||}^2-(m-\mathcal{T})^2-i\epsilon\right\}}
        -e^{-is\left\{p_{||}^2-\left(m+\sigma\mathcal{T}\right)^2-i\epsilon\right\}}
\right.
\nonumber \\
   &&\left.+\sum_{l=0}^{\infty}(2-\delta_{l0})
   e^{-is
    \left\{p_{||}^2-\left(\sqrt{m^2+2leB}
              +\sigma\mathcal{T}\right)^2-i\epsilon
    \right\}
   }\right\}
\label{poteff3}
\eea
where we are using the notation $p_{||}^2=p_0p^0+p_3p^3$.

Note that in (\ref{poteff3}) we added and subtracted the term $e^{-is\left\{p_{||}^2-\left(m+\sigma\mathcal{T}\right)^2-i\epsilon\right\}}$, and that the factor $(2-\delta_{l0})$ has been conveniently introduced despite once $\mathcal{T}$ is present,  there is no spin degeneracy for higher Landau levels.

Working in Euclidean space, we now perform in (\ref{poteff3}) the integration over
$p_{||}$ and the summation over $\sigma$, obtaining

\bea
   \Omega_{vac}^W&=& -\frac{|eB|}{2(2\pi)^2}
   \int_0^\infty
   \frac{ds}{s^2}e^{-s\left(m^2+\mathcal{T}^2\right)}
    \Bigg\{-\sinh(2s\mathcal{T}m)
\nonumber \\
  &&\left.+    \sum_{l=0}^{\infty}(2-\delta_{l0})
   e^{-2sleB}
   \cosh\left(2s\mathcal{T}\sqrt{m^2+2leB}\right)
     \right\}.
\label{poteff7}
\eea
By using the identity \cite{Davies,Gradshteyn}
\bea
  \cosh(x)=\frac{\sqrt{\pi}}{2\pi i}
        \int_{\gamma-i\infty}^{\gamma+i\infty}
           \frac{dt}{t^{\frac{1}{2}}}e^{t+\frac{x^2}{4t}}
\eea
with $\gamma$ a positive number, we can easily perform the sum over LL's
to get
\bea
   \Omega_{vac}^W&=&
   -\frac{|eB|}{2(2\pi)^2}
   \int_0^\infty
   \frac{ds}{s^2}
   e^{-s\left(m^2+\mathcal{T}^2\right)}
   {\Bigg\{ }
   -\sinh(2s\mathcal{T}m)
\nonumber \\
&&+
   \frac{\sqrt{\pi}}{2\pi i}
        \int_{\gamma-i\infty}^{\gamma+i\infty}
           \frac{dt}{t^{\frac{1}{2}}}e^{t+\frac{s^2\mathcal{T}^2m^2}{t}}
     \coth\left[s\left(1-\frac{s\mathcal{T}^2}{t}\right)eB\right]
    \Bigg\}
 \label{poteff85}
\eea
This particular form of the effective potential allows
us to isolate all divergences by making a Taylor expansion in powers
of ($eBs$) and ($\mathcal{T}$) up to order
$\mathcal{O}((eB)^4, \mathcal{T}^4)$, it is
\bea
   \Omega_{vac}^W&\approx& -\frac{1}{2(2\pi)^2}
   \int_0^\infty
   \frac{ds}{s^3}
   e^{-s m^2}\left(1-s\mathcal{T}^2+\frac{1}{2}s^2\mathcal{T}^4\right)
    |eB|s
\nonumber \\
    &&\times
   \left\{-2sm\mathcal{T}+
   \frac{\sqrt{\pi}}{2\pi i}
        \int_{\gamma-i\infty}^{\gamma+i\infty}
           \frac{dt}{t^{\frac{1}{2}}}e^{t}
     \left(1+\frac{s^2\mathcal{T}^2m^2}{t}+\frac{s^4\mathcal{T}^4m^4}{2t^2}\right)\right.
\nonumber \\
&&\times\left[\left(
  \frac{1}{eB s}+\frac{eB s}{3}-\frac{(eBs)^3}{45}\right)
 +\left(\frac{1}{(eBs)^2}-\frac{1}{3}+\frac{(eBs)^2}{15}
        -\frac{2 (eBs)^4}{189}
   \right) \frac{s\mathcal{T}^2eBs}{t}
\right..
\nonumber \\
 &&\left.\left.+\left(\frac{1}{(eBs)^3}-\frac{eB s}{15}+\frac{4 (eBs)^3}{189}\right) \frac{(s\mathcal{T}^2)^2(eBs)^2}{t^2}
     \right]\right\}
\eea
Hence, the divergent terms are
\bea
   \Omega_{div}^W&\approx&- \frac{1}{2(2\pi)^2}
   \int_0^\infty
   \frac{ds}{s^3}
   e^{-s m^2}\Bigg\{-2s^2m\mathcal{T}eB+
\nonumber \\
   &+&
   \frac{\sqrt{\pi}}{2\pi i}
        \int_{\gamma-i\infty}^{\gamma+i\infty}
           \frac{dt}{t^{\frac{1}{2}}}e^{t}
   \left[1+s \left(\frac{1}{t}-1\right)\mathcal{T}^2+
   s^2\left(\frac{(eB)^2}{3}+\frac{m^2
     \mathcal{T}^2}{t}+\frac{\mathcal{T}^4}{t^2}-\frac{\mathcal{T}^4}{t}+\frac{\mathcal{T}^4}{2}\right)\right]
   \Bigg\}
\eea
Integrating over $t$ term by term, we arrive at
\bea
    \Omega_{div}^W
    &=&  -\frac{1}{8\pi^2}
   \int_0^\infty
   \frac{ds}{s^3}
   e^{-s m^2}
     \left\{1-2ms^2\mathcal{T}eB+s\mathcal{T}^2+2s^2 m^2\mathcal{T}^2
       - \frac{1}{6}s^2\mathcal{T}^4+\frac{(eBs)^2}{3}
     \right\}
\label{poteff8}
\eea
where we used the identity~\cite{Davies}
\bea
  \frac{\sqrt{\pi}}{2\pi i}
        \int_{\gamma-i\infty}^{\gamma+i\infty}
           \frac{dt}{t^{\frac{1}{2}+n}}e^{t}=\frac{\sqrt{\pi}}{\Gamma\left(\frac{1}{2}+n\right)}.
\eea
Note that in (\ref{poteff8}) there are several $\mathcal{T}$-dependent divergences, similarly to what we also found in the strong-field limit. As in the $\mathcal{T}=0$ case \cite{Proper-Time}, where a vacuum term depending on the renormalized mass was subtracted, now we subtract the new magnetized vacuum term, which also depends on the particle AMM.
At  $\mathcal{T}=0$, the divergent term (\ref{poteff8}) reduces to that of Ref. \cite{Proper-Time}. The last term in the RHS of (\ref{poteff8}) is the usual divergent
term proportional to $B^2$ appearing at $T, \mu=0$ and that can be absorbed in the magnetic field
renormalization function $Z_3$  as is usually done
in the  $\mathcal{T}=0$ case.

Then, following the Schwinger procedure of renormalization, we
subtract \Eq{poteff8} to \Eq{poteff7}, to obtain the renormalized effective potential given by
\bea
   \Omega_{vac}^{W(R)}(B,0,0)&=&
   -\frac{1}{2(2\pi)^2}
   \int_0^\infty
   \frac{ds}{s^3}
   e^{-s\left(m^2+\mathcal{T}^2\right)}
   {\Bigg\{ }
   -|eB|s\sinh(2s\mathcal{T}m)
\nonumber \\
&+&
   \frac{\sqrt{\pi}}{2\pi i}
        \int_{\gamma-i\infty}^{\gamma+i\infty}
           \frac{dt}{t^{\frac{1}{2}}}e^{t+\frac{s^2\mathcal{T}^2m^2}{t}}
    |eB|s \coth\left(s\left(1-\frac{s\mathcal{T}^2}{t}\right)eB\right)
\nonumber \\
   &-&      \left(1+s\mathcal{T}^2+2s^2 m^2\mathcal{T}^2
       -2ms^2\mathcal{T}eB- \frac{1}{6}s^2\mathcal{T}^4+\frac{(eBs)^2}{3}
     \right)
   \Bigg\}
\label{poteff10}
\eea
Now $e$ and $B$ have their renormalized values that have absorbed the
divergent contribution $\Omega_{div}$. In the renormalization procedure we are following, we consider that $\mathcal{T}$ is a renormalized parameter, as it was considered in the strong-field case in Appendix A, and for the mass in the $\mathcal{T}=0$ case \cite{Proper-Time}. Then, the extracted divergency is considered as part of the renormalization of the vacuum energy that now also depends on the AMM.

In the weak-field expansion, the leading terms of the renormalized potential (\ref{poteff10}), taken  up to $\mathcal{O}((eB)^4)$ order and $\alpha^2$ correction, are
\bea
   \Omega_{vac}^{W(R)}(B,0,0)=\frac{1}{2(2\pi)^2}
    \left\{\frac{(eB)^4}{45m^4}
         +\frac{(eB)^2\mathcal{T}^2}{3m^2} \right\}
\label{poteffRen}
\eea

\subsection{Finite density at zero temperature}

The finite density contribution to the thermodynamical potential in the weak-field approximation is given by
\bea
   \Omega^{W}_\mu=-\frac{|eB|}{(2\pi)^2}
   \int dp_{3}
 \left\{
   (\mu-\varepsilon_{++0})\theta(\mu-\varepsilon_{++0})
   +
   \sum_{\sigma=\pm1}
   \sum_{l=1}^{\infty}
          (\mu-\varepsilon_{+\sigma l})\theta(\mu-\varepsilon_{+\sigma l})
   \right\}
\eea

In the weak-field limit, the sum over all LL's can be easily done if
we rewrite the above equation as follows
\bea
   \Omega^{W}_\mu&=&-\frac{1}{2}\sum_{\sigma=\pm1}\frac{|eB|}{(2\pi)^2}
   \int dp_{3}
  \Bigg\{
   (\mu-\varepsilon_{++0})\theta(\mu-\varepsilon_{++0})
   -(\mu-\varepsilon_{+\sigma0})\theta(\mu-\varepsilon_{+\sigma0})
\nonumber\\
   &&\hspace{3cm}\left.+\sum_{l=0}^{\infty}(2-\delta_{l0})
          (\mu-\varepsilon_{\eta\sigma l})\theta(\mu-\varepsilon_{\eta\sigma l})
   \right\}
   \label{TP-1}
\eea

Taking into account that the separation between LL's is given by
$\sqrt{2eB}$, in the weak-field limit the LL's are so close that we
can consider that they form a continuos
distribution. Therefore, this allows us to  use the
  Euler-Maclaurin formula~\cite{EULERMAC}
\bea
   \frac{|eB|}{2}\sum_{l=0}^{\infty}(2-\delta_{l0})f(2eBl)&\approx&\int_0^\infty
   dx
   f(x)+\frac{|eB|}{2}f(\infty)
\nonumber \\&&
+\sum_{k=1}^{\infty}\frac{(eB)^{2k}}{(2k)!}B_{2k}\left\{f^{(2k-1)}(\infty)-f^{(2k-1)}(0)\right\}
\eea
where  $x$ is a continuous variable and $B_{2k}$ are the
  Bernoulli numbers
with $B_0=1$, $B_1=-\frac{1}{2}$ and $B_2=\frac{1}{6}$.

Thus, (\ref{TP-1}) up to $\mathcal{O}((eB)^2)$ can be expressed as
\bea
  \Omega^{W}_\mu&=&-\frac{1}{(2\pi)^2}
   \sum_{\sigma=\pm1}\Bigg\{
      \int \frac{d^3p}{(2\pi)^3}
          (\mu-\varepsilon_{+\sigma p_\perp})\theta(\mu-\varepsilon_{+\sigma p_\perp})
\nonumber \\
   &&\hspace{2cm}+\frac{|eB|}{2}
   \int dp_{3}
   \Big[(\mu-\varepsilon_{++0})\theta(\mu-\varepsilon_{++0})
   -(\mu-\varepsilon_{+\sigma0})\theta(\mu-\varepsilon_{+\sigma0})\Big]
\nonumber\\
   &&\hspace{2cm}+\frac{|eB|^2}{12}\int\ dp_3
    \frac{m+\sigma \mathcal{T}}{m\varepsilon_{+\sigma 0} }
    \theta(\mu-\varepsilon_{+\sigma0})
  \Bigg\}
\label{TP-1a}
\eea
where
\bea
  \varepsilon_{\eta\sigma p_\perp}
   \equiv  \eta\sqrt{p_{3}^2+\left(\sqrt{m^2+p_\perp^2}+\sigma\mathcal{T}\right)^2}
\label{varepsilondef-2}
\eea
and we replaced $x$ by $p_\perp^2$ in the first term of \Eq{TP-1a}.

Performing the integration over all momenta and Taylor
expanding in powers of $\mathcal{T}$, the leading
contribution is given by
\bea\label{TP-mu}
  \Omega^W_\mu
    &=&-\frac{1}{8\pi^2}
 \Bigg(\frac{2}{3}\mu p_F^3-m^2\mu p_F
       +m^4\ln\left(\frac{p_F+\mu}{m}\right)\Bigg)\theta(\mu-m)
\nonumber \\
      &&-\frac{\mathcal{T}^2}{4\pi^2}\left(\mu p_F+m^2\ln\left(\frac{p_F+\mu}{m}\right) \right)\theta(\mu-m)
+\frac{|eB|m\mathcal{T}}{2\pi^2}\ln\left(\frac{p_F+\mu}{m}\right)\theta(\mu-m)
\nonumber \\
   &&
    -\frac{|eB|^2}{12\pi^2}\ln\left(\frac{p_F+\mu}{m}\right)\theta(\mu-m)
    +\mathcal{O}(|eB|^4)
\eea
where $p_F\equiv\sqrt{\mu^2-m^2}$.

\subsection{Finite-temperature}

The finite temperature  thermodynamical potential reads
\bea
   \Omega^{W}_\beta
   =-\frac{|eB|}{(2\pi)^2\beta}
   \int dp_{3}\sum_{\eta=\pm1}
   \left\{
     \ln\left(1+e^{-\beta |\varepsilon_{\eta+0}+\mu|}\right)
   +\sum_{\sigma=\pm1}
   \sum_{l=1}^{\infty}
         \ln\left(1+e^{-\beta |\varepsilon_{\eta\sigma l}+\mu|}\right)
   \right\}
\nonumber \\
\label{poteffdens2}
\eea
Once we perform the sum over $\eta$ we get
\bea
   \Omega^{W}_\beta
   &=&-\frac{|eB|}{(2\pi)^2\beta}
   \int dp_{3}
     \ln\left(1+e^{-\beta |\varepsilon_{++0}+\mu|}\right)
        \left(1+e^{-\beta |\varepsilon_{++0}-\mu|}\right)
\nonumber \\
   &&-\frac{|eB|}{(2\pi)^2\beta}
   \int dp_{3}\sum_{\sigma=\pm1}
   \sum_{l=1}^{\infty}
         \ln\left(1+e^{-\beta |\varepsilon_{+\sigma l}+\mu|}\right)
            \left(1+e^{-\beta |\varepsilon_{+\sigma l}-\mu|}\right)
\nonumber \\
\label{poteffdens2c}
\eea

\end{document}